\documentclass[]{aa}

\usepackage{graphicx}
\usepackage{times}
\usepackage{amsmath}
\usepackage{color}
\usepackage{graphics}
\usepackage{subfigure}
\usepackage{natbib}
\usepackage[flushleft]{threeparttable}

\bibpunct{(}{)}{;}{a}{}{,}

\begin{document}

   \title{The transition from eruptive to confined flares in the same active region}

   \subtitle{}

   \author{F.~P.~Zuccarello\inst{1}\inst{,2}
	\and
	R.~Chandra\inst{3}
	\and
	B.~Schmieder\inst{2}
	\and
	G. Aulanier\inst{2}
	\and
	R.~Joshi\inst{3}}

\institute{Centre for mathematical Plasma Astrophysics, Department of Mathematics, KU Leuven, Celestijnenlaan 200B, B-3001 Leuven, Belgium\\
           \email{Francesco.Zuccarello@kuleuven.be}
           \and
           LESIA, Observatoire de Paris, PSL Research University, CNRS, Sorbonne Universit\'{e}s, 
           UPMC Univ. Paris 06, Univ. Paris-Diderot, Sorbonne Paris Cité, 5 place Jules Janssen, F-92195 Meudon, France
           \and
           Department of Physics, DSB Campus, Kumaun University, Nainital 263001, India }

\date{Received / Accepted}

\titlerunning{Eruptive and confined flares in the same active region}
\authorrunning {F.~P.~Zuccarello et al.}

% \abstract{}{}{}{}{} 
% 5 {} token are mandatory
 
  \abstract
{%Context
Solar flares are sudden and violent releases of magnetic energy in the solar atmosphere that can be divided in eruptive flares, when plasma is ejected from the solar atmosphere, resulting in a coronal mass ejection (CME), and confined flares when no CME is associated with the flare. 
}
{%Aims
We present a case-study showing the evolution of key topological structures, such as spines and fans  which may determine the eruptive versus non-eruptive behavior of the  series of eruptive flares, followed by confined flares, which are all originating from the same site.
}
{%Methods
To study the connectivity of the different flux domains and their evolution, we compute a potential magnetic field model of the active region. Quasi-separatrix layers are retrieved from the magnetic field extrapolation.
}
{%Results
The change of behavior  of the flares from one day to the next ---eruptive to confined--- can be attributed to the change of orientation of the magnetic field below the fan with respect to the orientation of the overlaying spine, rather than an overall change in the stability of the large scale field. 
}
{%Conclusions
Flares tend to be more-and-more confined when the field that supports the filament and the overlying field  gradually become less-and-less anti-parallel, as a direct result of changes in the photospheric flux distribution, being themselves driven by continuous shearing motions of the different magnetic flux concentrations.
}


   \maketitle
%
%-------------------------------------------------------------------

\section{Introduction}

\begin{figure*}
\begin{center}
\subfigure[]{
\includegraphics[width=.32\textwidth,viewport= 0 0 512 512,clip]{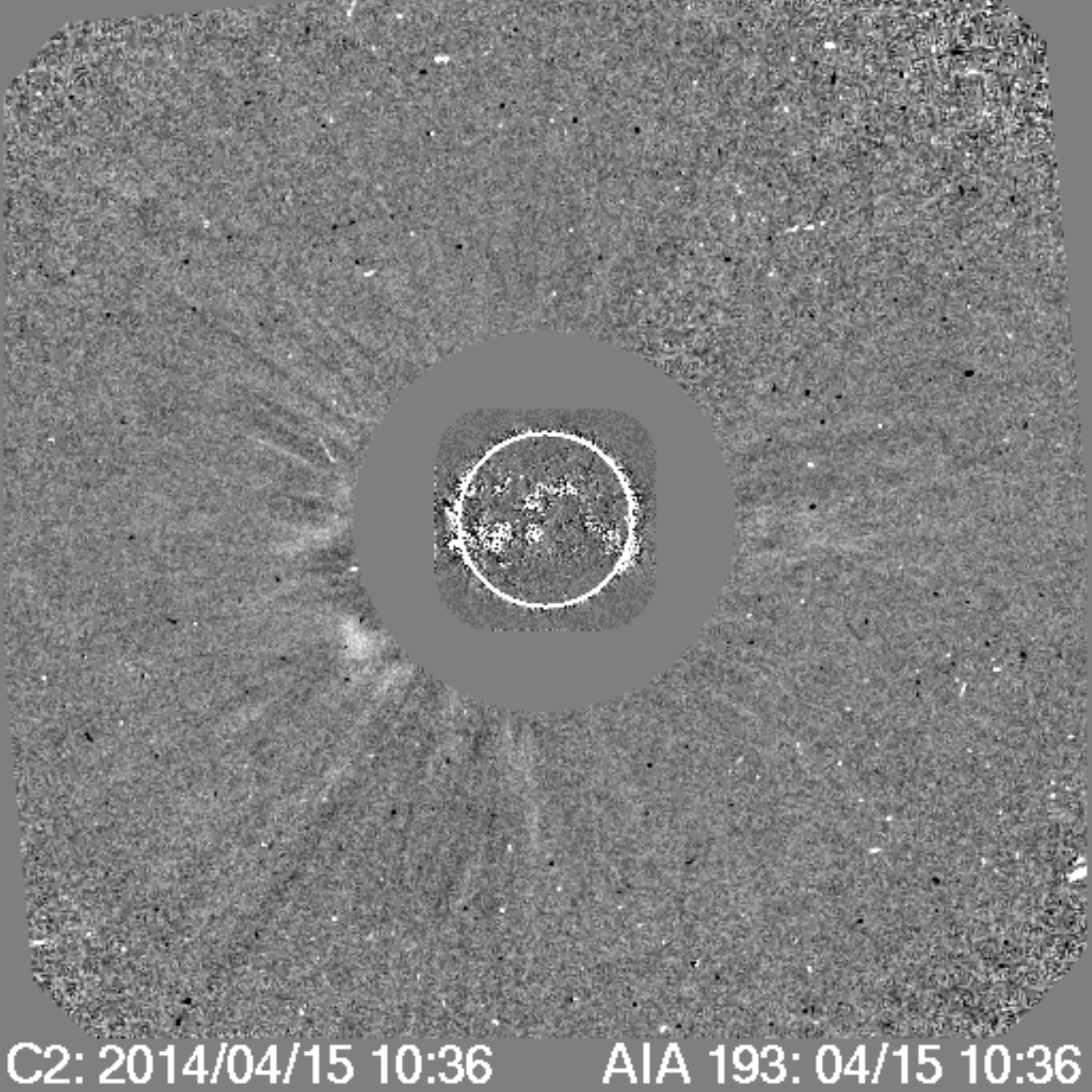}\label{Fig:OBS-CMEa}}
\subfigure[]{
\includegraphics[width=.32\textwidth,viewport= 0 0 512 512,clip]{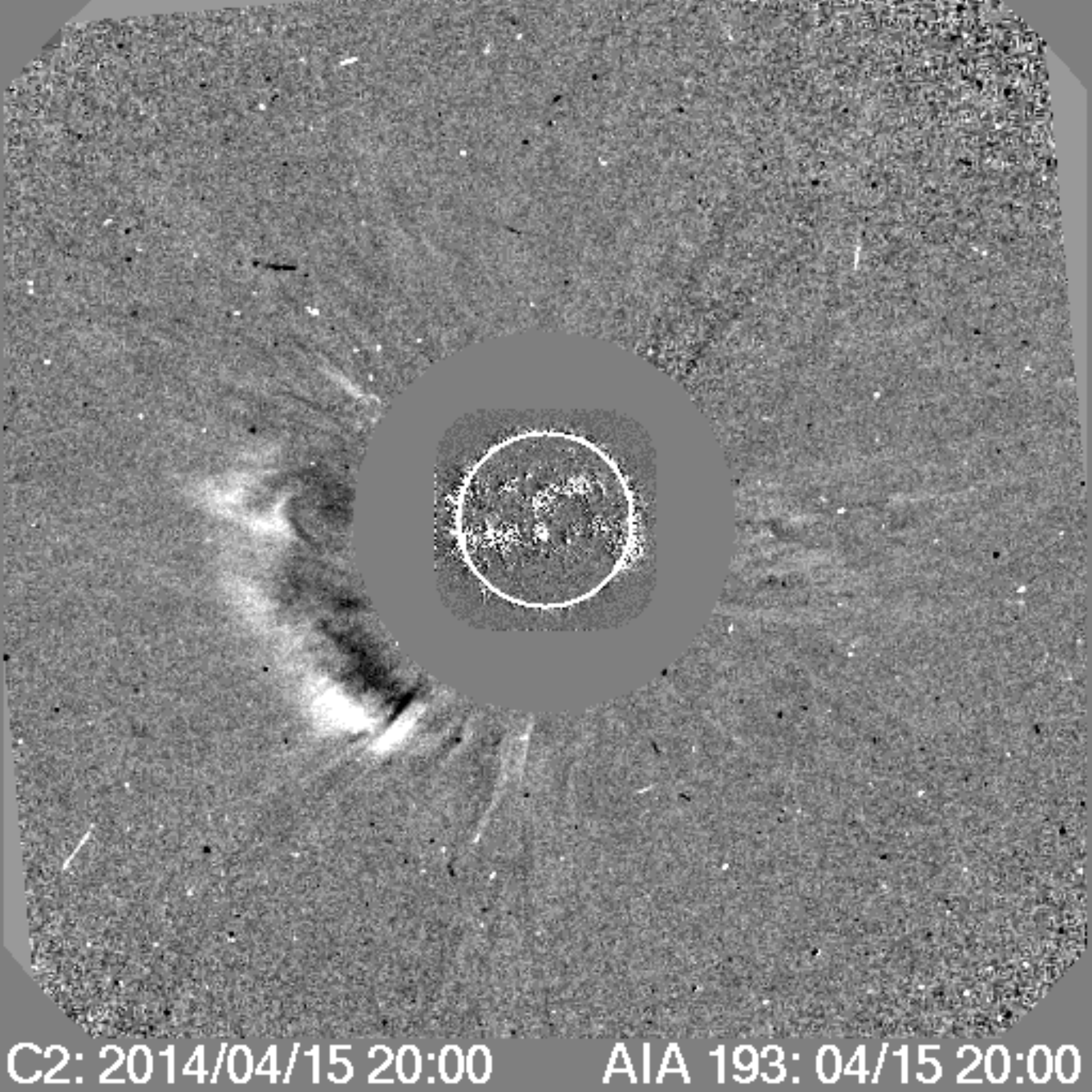}\label{Fig:OBS-CMEb}}
\subfigure[]{
\includegraphics[width=.32\textwidth,viewport= 0 0 512 512,clip]{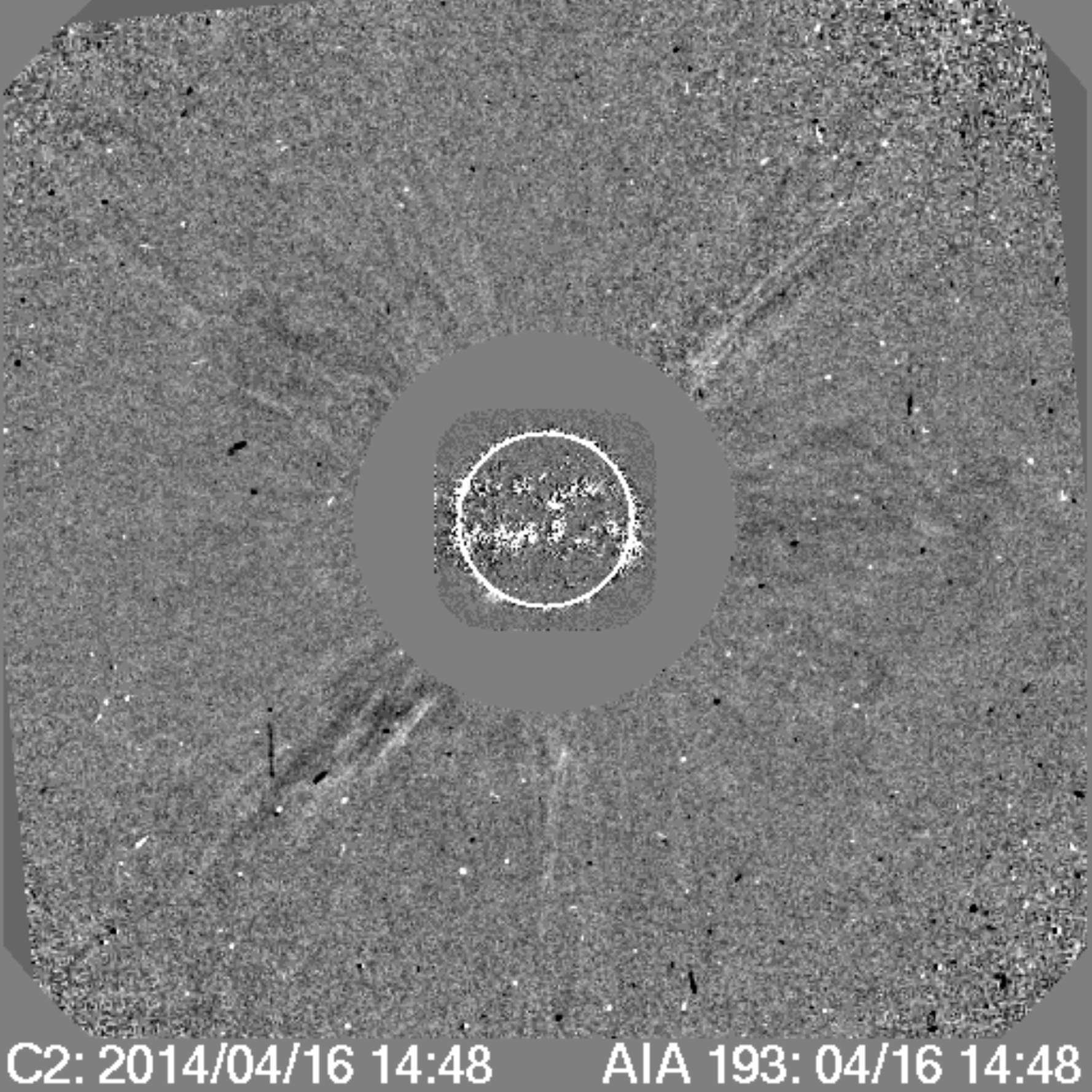}\label{Fig:OBS-CMEc}}
\caption{ SOHO/LASCO running difference images showing the associated CMEs for the eruptive flares. The  last panel shows that no CME could be detected for the April 16 failed eruption.
\label{Fig:OBS-CMEs}}
\end{center}
\end{figure*}

\begin{figure*}
\begin{center}
\subfigure[]{
\includegraphics[width=.32\textwidth,viewport= 2 9 375 245,clip]{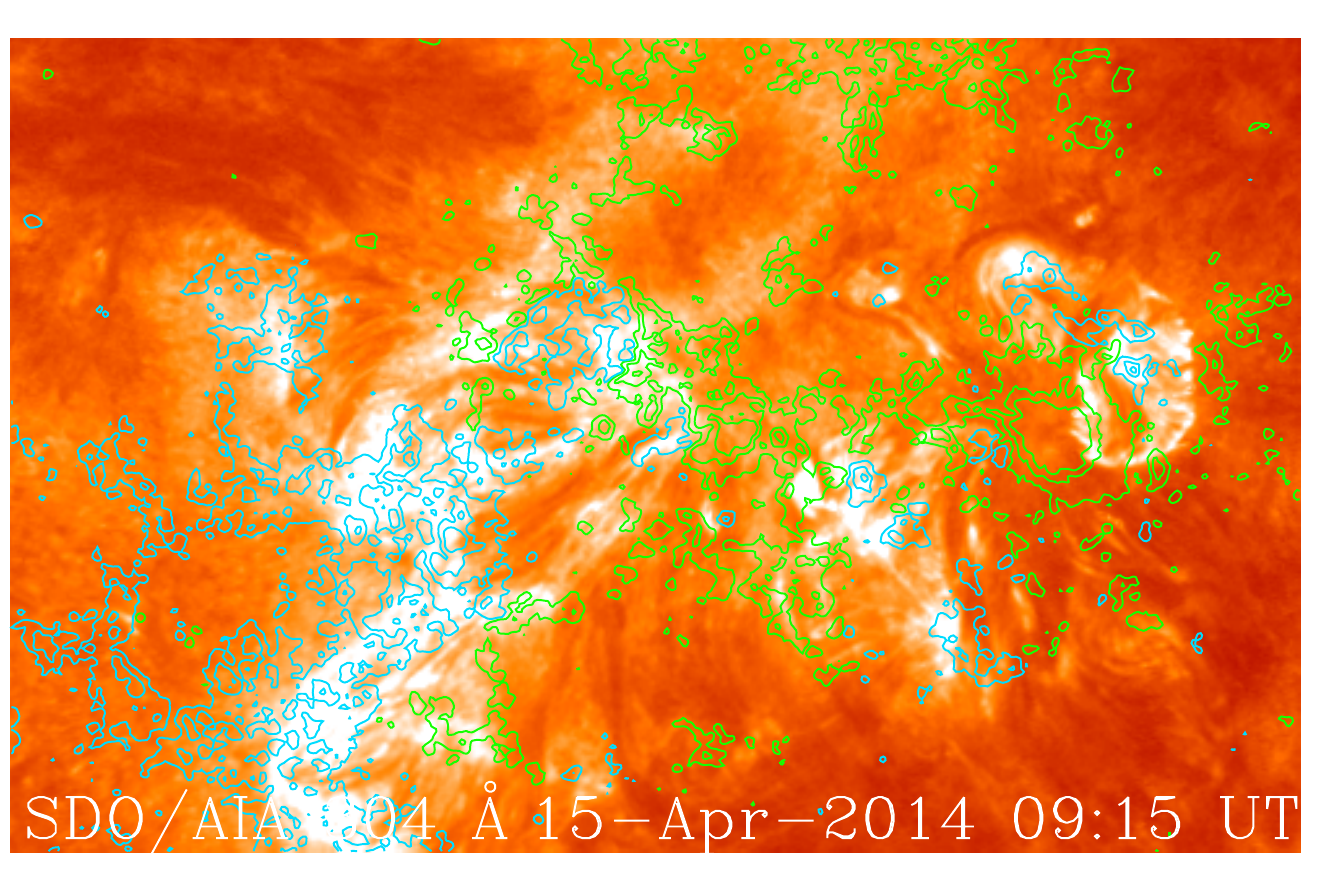}\label{Fig:OBS-15-04a}}
\subfigure[]{
\includegraphics[width=.32\textwidth,viewport= 2 2 375 238,clip]{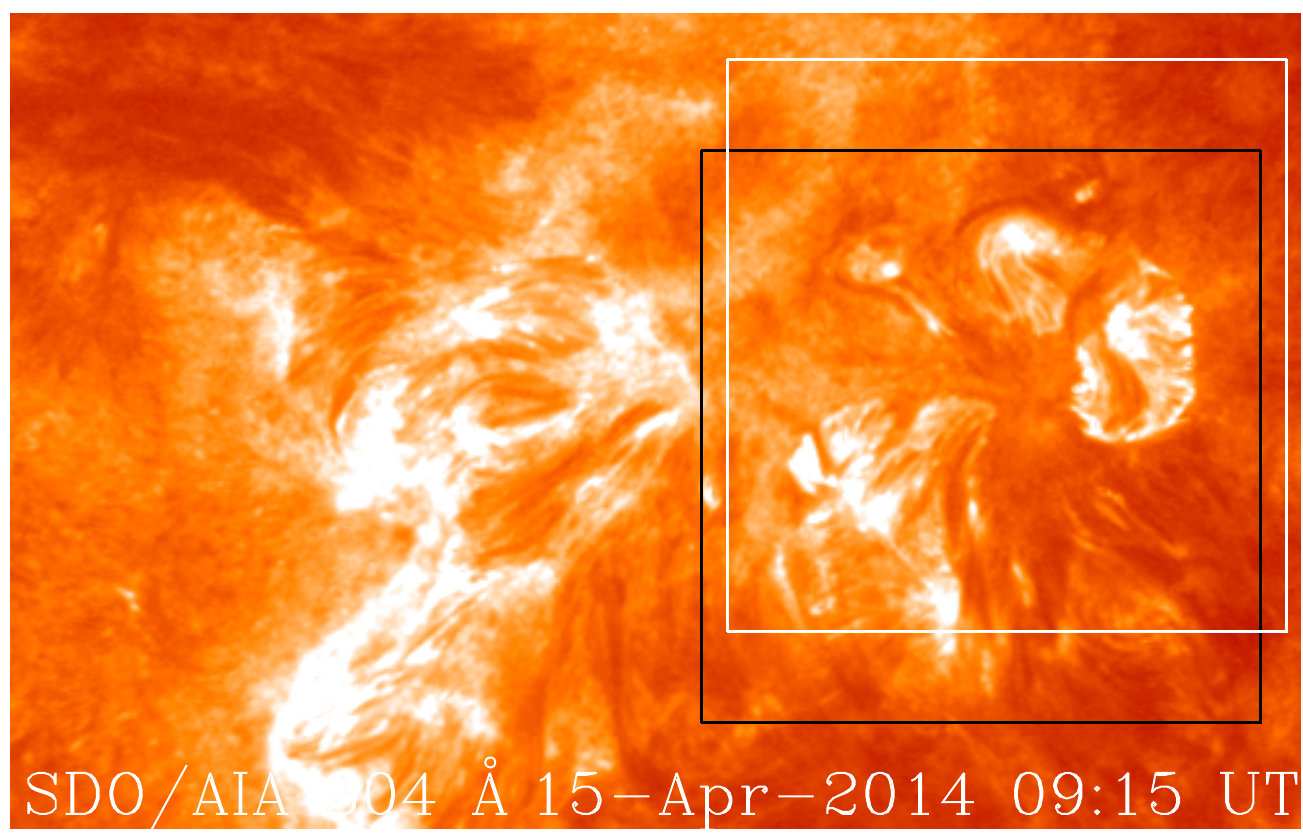}\label{Fig:OBS-15-04b}}
\subfigure[]{
\includegraphics[width=.32\textwidth,viewport= 7 9 380 245,clip]{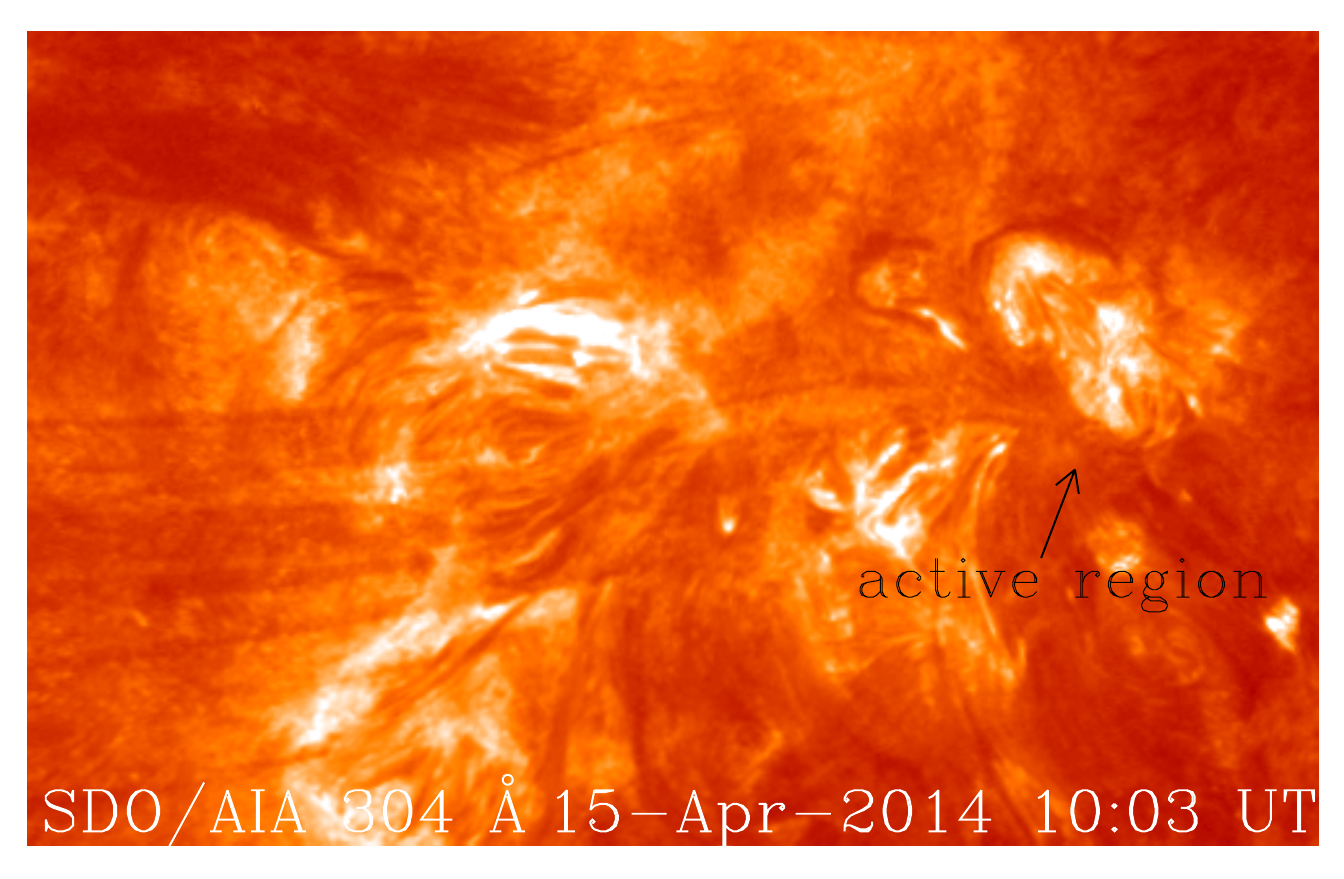}\label{Fig:OBS-15-04c}}
\subfigure[]{
\includegraphics[width=.32\textwidth,viewport= 7 9 380 245,clip]{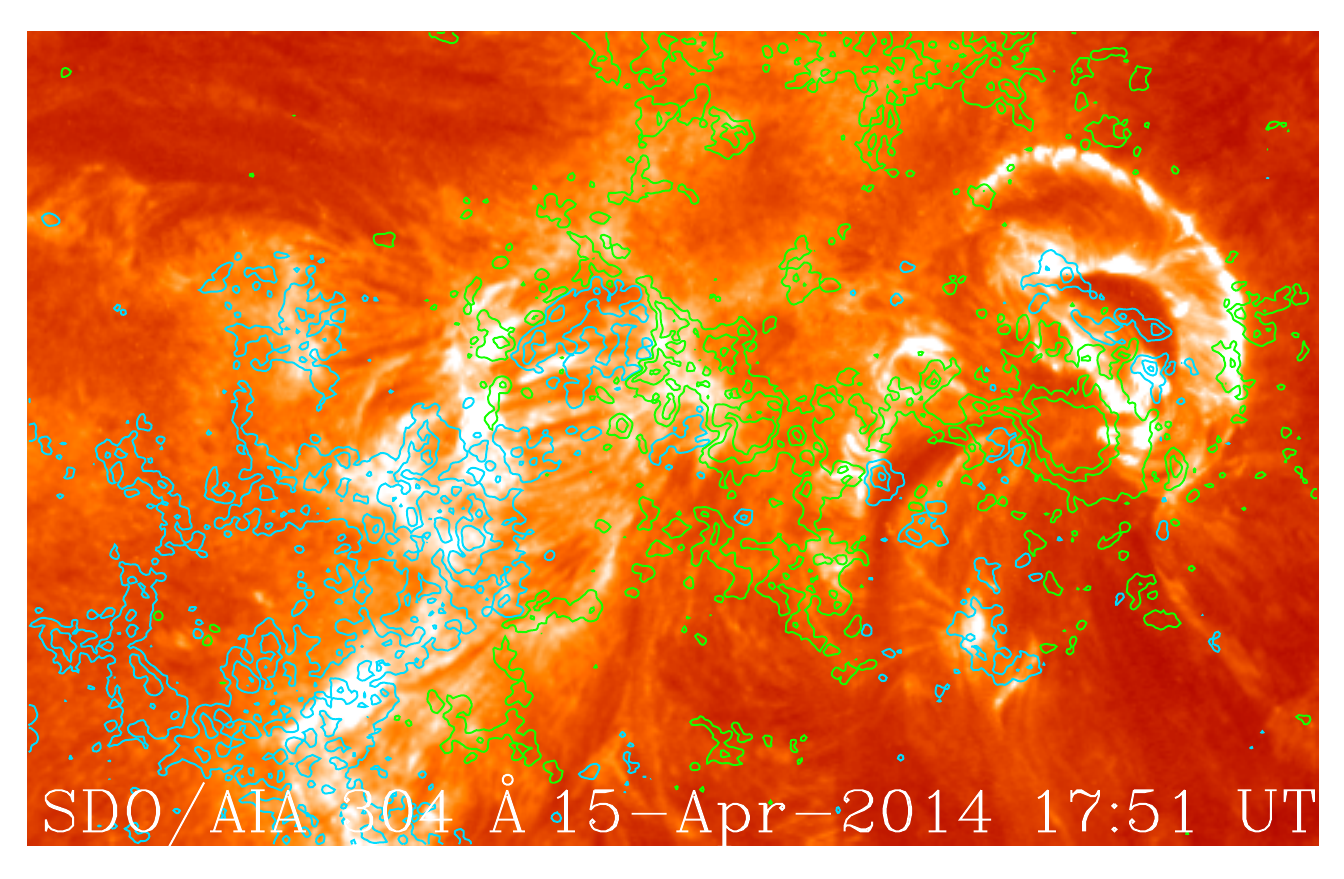}\label{Fig:OBS-15-04d}}
\subfigure[]{
\includegraphics[width=.32\textwidth,viewport= 2 2 375 238,clip]{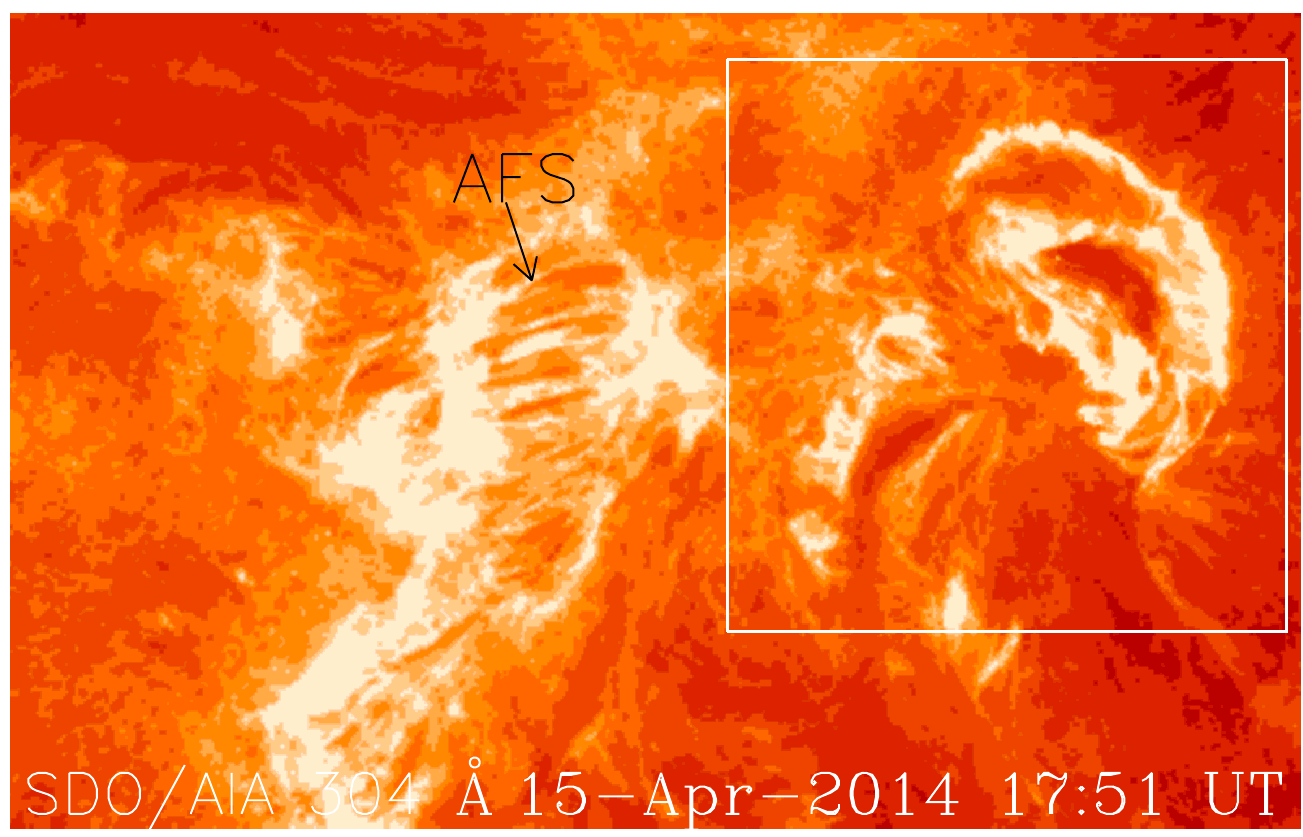}\label{Fig:OBS-15-04e}}
\subfigure[]{
\includegraphics[width=.32\textwidth,viewport= 2 9 375 245,clip]{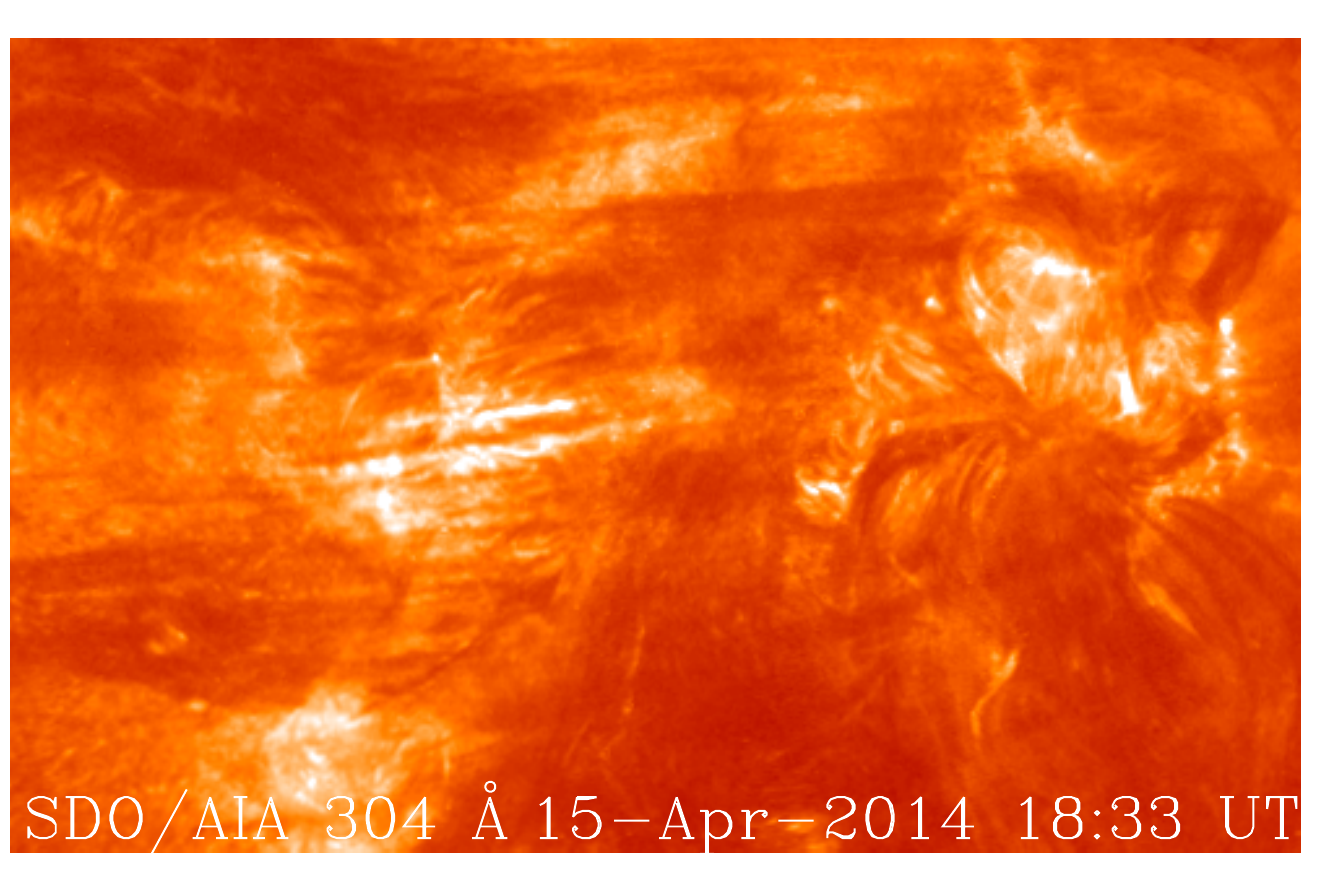}\label{Fig:OBS-15-04f}}
\subfigure[]{
\includegraphics[width=.32\textwidth,viewport= 7 9 380 245,clip]{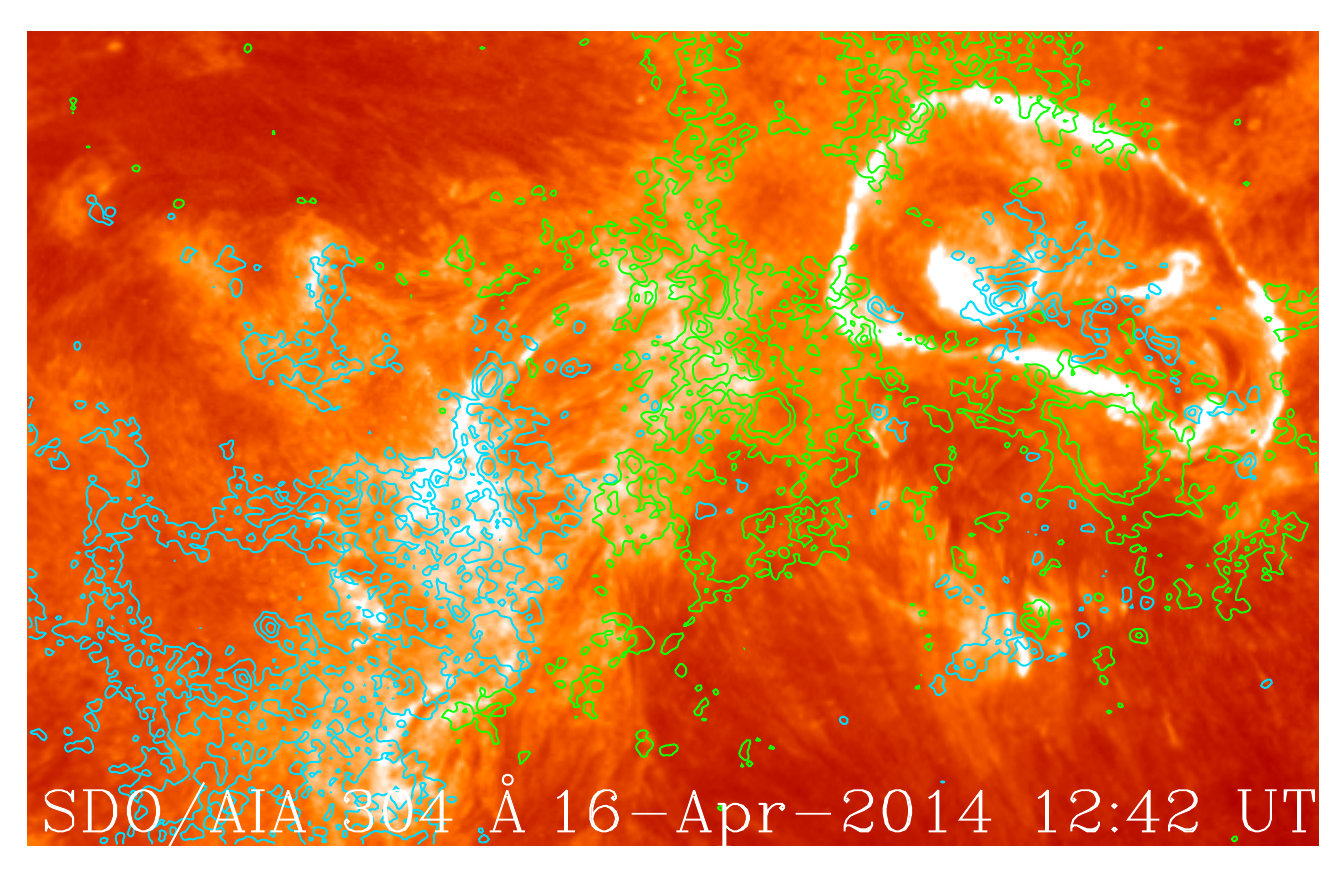}\label{Fig:OBS-15-04g}}
\subfigure[]{
\includegraphics[width=.32\textwidth,viewport= 2 2 375 238,clip]{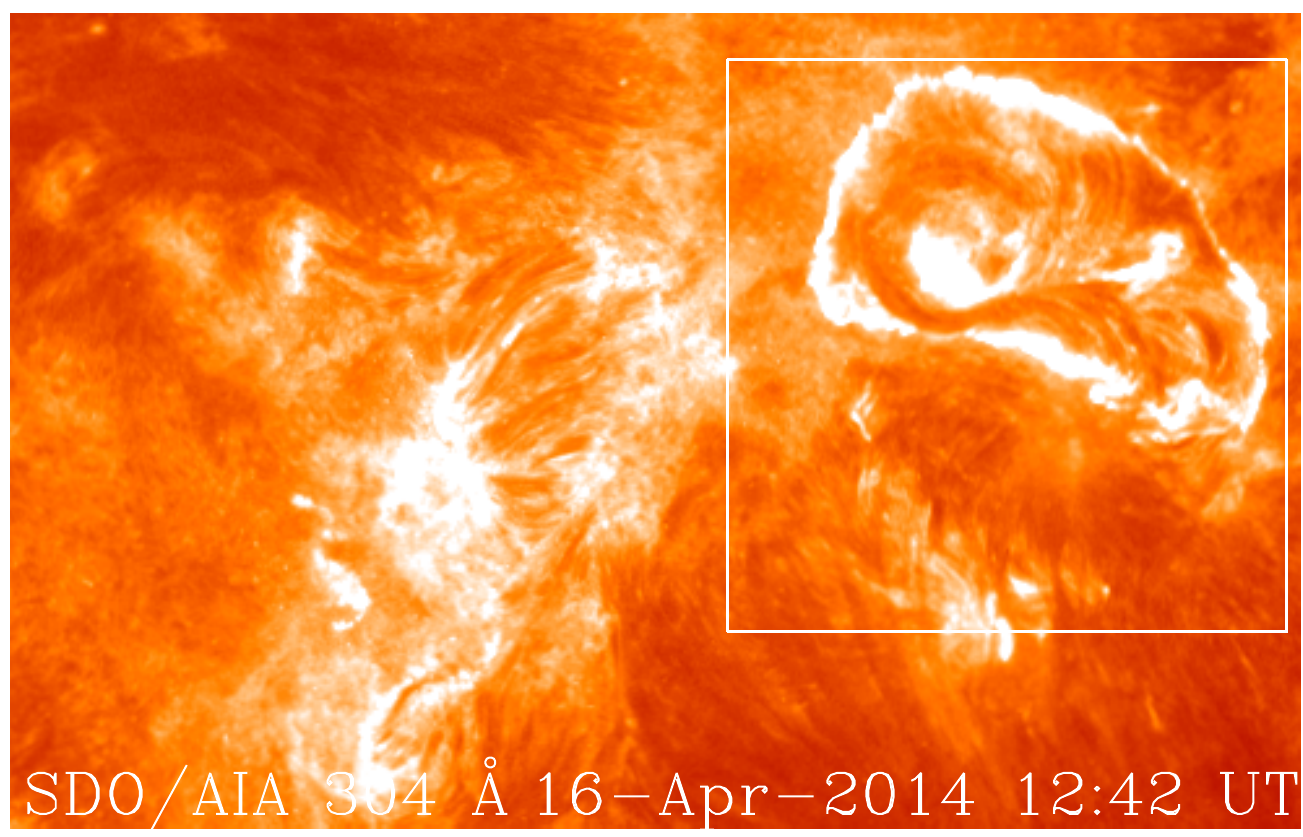}\label{Fig:OBS-15-04h}}
\subfigure[]{
\includegraphics[width=.32\textwidth,viewport= 7 9 380 245,clip]{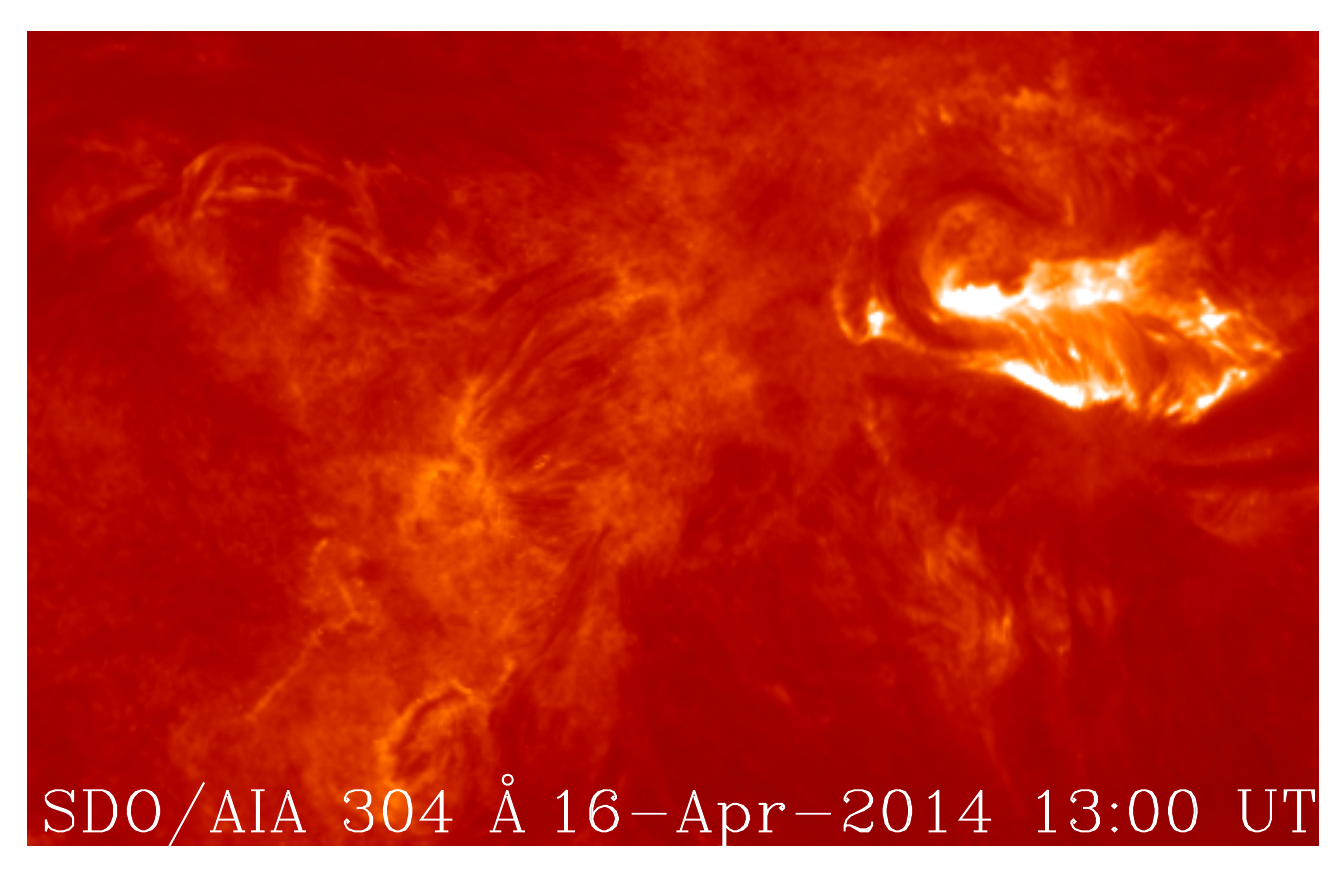}\label{Fig:OBS-15-04i}}
\caption{Two examples of the C-class eruptive flares in the AR 12035 on 2014 April 15  in AIA 304 $\AA$ (two top rows) and one example of confined flare on 2014 April 16 (bottom row). The FOV is [300 $\times$ 200] arcseconds and the active region corresponds to the bright region in the right of the images indicated by an arrow in panel $c$.  The arch filament system (AFS) discussed in Section~\ref{Sect:MF} is indicated by an arrow in panel $e$.
%First column represents the enlarged field-of-view of the region enclosed by the black box in the second column. 
The left column images are overlaid by HMI LOS magnetic field contours. Green/cyan contours (levels: $\pm$ 100, $\pm$ 500, $\pm$ 900) represent positive/negative polarity respectively. The black square in panel $b$ indicates the approximate FOV of Figure~\ref{Fig:filament}, while the white square indicates the approximate FOV of Figure~\ref{Fig:T-QSL}. The temporal evolution is shown in Movie~1 available online.
\label{Fig:OBS-15-04}}
\end{center}
\end{figure*}

\begin{figure*}%[!t]
	\centering
    \includegraphics[width=.99\textwidth,viewport= 15 85 719 527,clip] {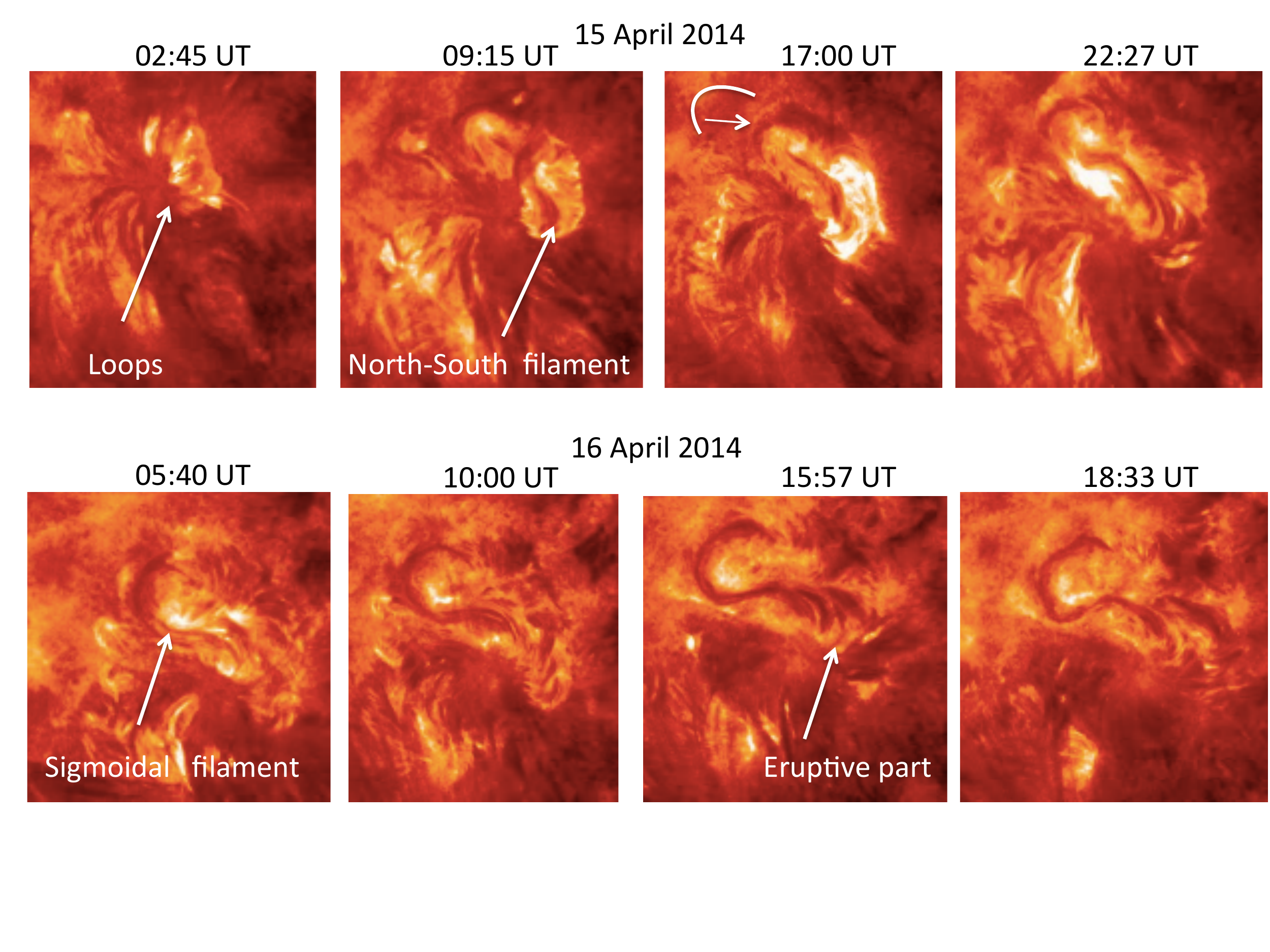}
     \caption{Evolution of the filaments in the north part of AR 12035 between April 15 and 16  in AIA 304 $\AA$.  Note the north south filament on April 15 at 09:15 UT and the upside down U filament  at 17:00 UT.  A    sigmoidal filament   is formed on April 16. The FOV corresponds to the black box in Figure \ref{Fig:OBS-15-04}. The temporal evolution is shown in Movie~2 available online.
\label{Fig:filament}}
\end{figure*}

Solar flares are one of the most energetic phenomena that occur in the solar atmosphere, and they can be divided into eruptive flares that are associated with coronal mass ejections, and confined flares \citep[see reviews of][]{Sch2015, Jan2015}. These latter are normally not associated with  a CME, and either  no filament 
is present at all \citep{Sch1997, Dal2015}  or the filament fails to erupt \citep{Tor2005,Guo2010a}. In addition to full and failed eruptions, there are cases where only a part of the filament is erupted, such events are defined as partial erupting events. Partial eruption may or may not be associated with a CME \citep{Gibson06, Liu08,Liu12,Tripathi13,Kliem14, Zhang14}. 

The most energetic flares are commonly eruptive \citep{Yas2005}, even though confined, non-eruptive X-class flares have been reported \citep{Tha2015, Sun2015, Har2016} as well as CMEs  with associated only C-class flares \citep{Rom2014, Chandra2016}.

The CSHKP model \citep{Carmichael64,Sturrock66,Hirayama74,Kopp76} and its extension in three-dimensions \citep{Aul2012,Jan2013,Jan2015} can explain several observational signatures of the fully (or failed) eruptive flares, such as the presence of X-ray sigmoids,  flare ribbons, and  brightening motions along the ribbons themselves. In particular, \cite{Sav2015,Sav2016} have shown that the flare ribbons often coincide with the photospheric signature of quasi-separatrix layers \citep[QSLs,][]{Dem1996}, i.e., thin layers characterized by a sharp gradient in the connectivity of the magnetic field. The brightening motions along the ribbons have been interpreted as the signatures of the slipping reconnection of the magnetic field lines through the QSL \citep{Aul2006, Jan2013, Dud2014a, Dud2016}. 

The morphology and evolution of  flare ribbons can also give information on the overall topology of the system. \cite{Mas2009} have shown that circular flare ribbons are associated with the presence of a null-point topology in the corona, while parallel ribbons moving  away from  each other have been interpreted as  an indication of quasi-separator reconnection occurring higher and higher in the corona \citep{Aul2012}. 

From a theoretical point of view, a key feature of the standard model for solar flares is the presence of an outward moving magnetic flux rope, i.e., a topological structure constituted by twisted magnetic field lines that wrap around an axial magnetic field line. Structures compatible with magnetic flux ropes have been observed both on an active region scale \citep{Can2010,Jing2010,Guo2010b,Gre2011,Sav2012,Gibb2014,Jiang2014} and on a larger scale in coronal cavities \citep[][Bak-Steslicka et al. 2016]{Gib2010,Rac2013, Gib2015}. The standard flare model requires a mechanism that triggers the onset of the flux rope eruption, resulting in the phenomenology observed during solar flares. 

Different triggering mechanisms have been proposed and discussed in the literature \citep[see][for a review]{Forbes2010,Chen2011,Aul2014,Fil2015,Sch2015}, but essentially the equilibrium of a magnetic flux rope embedded in an overlying magnetic field is determined by two competing effects: the outward-directed magnetic pressure between the flux rope and photosphere, and the inward-directed magnetic tension of the overlying field. 

In the torus instability or catastrophic loss of equilibrium model \citep{For1991,Kliem2006,Dem2010,Kliem2014} it  is the onset of an ideal magneto-hydrodynamic instability that leads to the disruption of this equilibrium, while in the breakout model \citep{Ant1999, Lyn2008, Zuc2008, Zuc2009,Karpen2012} it is the onset of a resistive instability.   

Assuming an overlying external field $B_{\text{ex}}$ that scales with the height $z$ from the photosphere as $B_{\text{ex}} \propto z^{-n}$, in the torus instability model the system becomes unstable when the apex of the axis of the magnetic flux rope reaches a critical height $z_{cr}$ where the decay index $n$ of the external overlying field $B_{\text{ex}}$ becomes larger than a critical value $n_{\text{cr}}$. The results of several MHD simulations place $n_{\text{cr}}$ in the range $[1.3-1.75]$ \citep{Tor2005,Tor2007, Fan2007,Ise2007,Aul2010, Kliem2013,Ama2014,Ino2015, Zuc2015, Zuc2016}.  Attempt to estimate the decay index at the onset of solar eruptions have also been made both using limb observations as well as stereoscopic observations \citep{Fil2001,Guo2010a, Fil2013,Zuc2014,McC2015}. These studies have found an `observed' critical decay index $n_{\text{cr}}^{\text{ob}}$ in the range $[1-1.1]$, with this discrepancy between models and observations partially due to the different location `where' the decay index is computed \citep{Zuc2016}.  

Contrary to the torus instability model that does not require any particular magnetic field topology, the breakout model requires a multi-flux distribution. The eruption begins when a resistive instability sets in at the so-called breakout current sheet that exists between the arcade that confines the flux rope and the overlying field \citep{Karpen2012}. This reconnection removes the confining flux by transferring it to the neighboring flux domains. As a result, the magnetic tension of the confining field decreases resulting in an eruption. For the breakout model to work two conditions must be satisfied: the presence of a null-point or quasi-separator in the corona, 
and the flux of the confining arcade must be larger than the flux of the overlying field. 
Due to the nature of the problem, i.e., evidence of reconnection occurring higher up in the corona, observational studies that clearly support the breakout model are quite rare \citep{Aulanier2000,Man2006,Chandra2009,Chen2016}.

Both models address the triggering of filament eruption, but what determines if a filament eruption results in a CME or in a failed eruption ?
Many questions have to be answered:
Does/how does the trigger mechanism affect the eruptive/failed behavior of the flare ?
How important is the magnetic environment of the active region ?
 
In order to address these questions and understand what causes confined or eruptive flares, we study a series of flares that occurred between 2014 April 15 and 16 in active region NOAA 12035 and that resulted in full eruptions on April 15 and in failed eruptions on April 16. The paper is organized as follows. In the next Section we present our observational datasets. The analysis of the magnetic topology of the active region is presented in Section 3. Finally, in Section 4 we discuss our results and conclude in Section 5.

\begin{figure}%[!t]
	\centering
    \includegraphics[width=.50\textwidth] {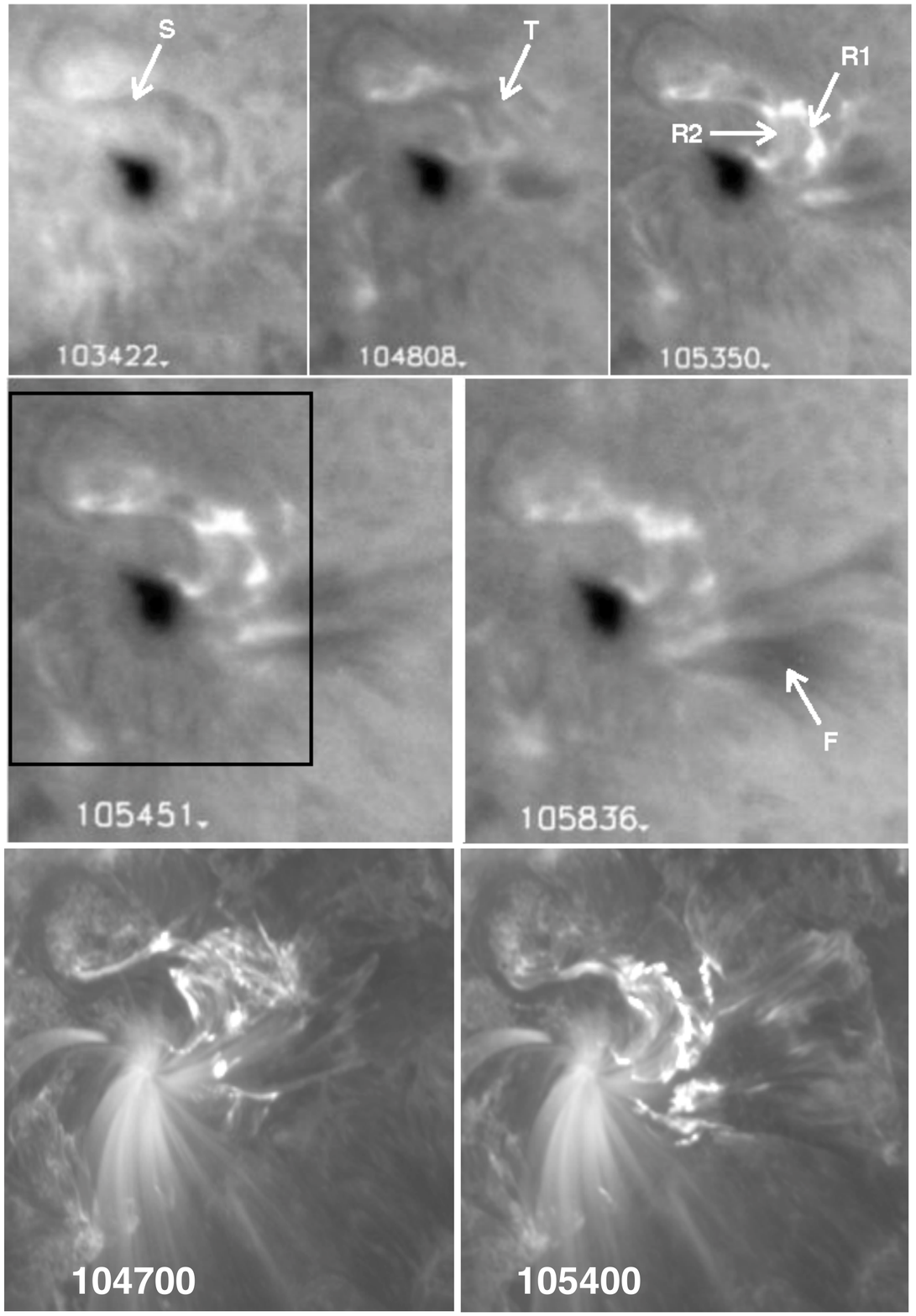}
     \caption{H$\alpha$ images of the failed filament eruption and of the flare on 16 April 2014 from 10:34 UT -- 10:58 UT  (from Nainital ARIES telescope, top rows). The box in the left middle image represents the FOV of the images in the top row. In the bottom row,  the corresponding images in AIA 171 \AA\  at 10:47UT  and 10:54 UT are shown. The FOV of the bottom panels is similar to the FOV of the middle panels. See text for more details on the S-shaped filament (S), the multiple threads (T), the two bright flare ribbons (R1, R2), and the black fan-like shapes (F).
\label{Fig:Halpha}}
\end{figure}

\begin{table}
\caption{Details of compact and eruptive flares. }
\label{recurrent}
\begin{center}
\begin{threeparttable}

%\begin{large}
\begin{tabular}{cccc}
\hline
\multicolumn{4}{|c|}{\bf 15 April 2014}\\
\hline
Number  &Flare onset &  Flare class & CME \\
of flare & (UT) & GOES& association\\
\hline
1 &  05:56& -- & No \\
2 &  06:15& -- & No\\
3 &  06:59& -- & No\\
\textbf{4} &  \textbf{09:15}& \textbf{C8.6} & \textbf{10:24}\\
5 &  12:34& C3.6  & 14:00\\ %erupt until 13:54 UT
6 &  14:37& -- & No\\
7 &  16:56& -- & 14:00\\
\textbf{8} &  \textbf{17:53}& \textbf{C7.3} & \textbf{18:48}\\
9 &  19:22& -- & No \\
10&  20:55& -- & No \\
11&  21:39& -- & No \\
12&  22:48& -- & No \\
13&  23:40& -- & No \\
\hline
\multicolumn{4}{|c|}{\bf 16 April 2014}\\
\hline
14&  01:10 & C1.9 & No\\
15&  02:42 &  --  & No\\
16&  03:20 &  --  & No\\
17&  03:48 &  --  & No\\
18&  05:02 &  --  & No\\
19&  06:37 & C1.8 & $X$ \\%
20&  07:14 &  --  & No\\
21&  08:36 & C5.2 & No\\
22&  09:20 & -- & No\\
\textbf{\underline{23}}& \textbf{\underline{10:42}} & \textbf{--} & \textbf{\underline{No}}\\
\textbf{24}& \textbf{12:42} & \textbf{C7.5 }& \textbf{No}\\
25&  17:30 & C2.0 & No\\
26&  19:54 & M1.0 & $X$ \\%CME\\
\hline
\end{tabular}
%\end{large}
\begin{tablenotes}
      \small
      \item \textbf{Note.} Flares indicated in boldface are shown in Figure~\ref{Fig:OBS-15-04}, while the flare observed in H$\alpha$ and also presented in Figure~\ref{Fig:Halpha} is underlined.  The '--' indicates small flares that are not reported by GOES,  '$X$' indicates CMEs visible in LASCO, but not associated with the filament activity/eruption.  
    \end{tablenotes}
  \end{threeparttable}
  \end{center}
\end{table}

\begin{figure*}
\begin{center}
\includegraphics[width=.99\textwidth]{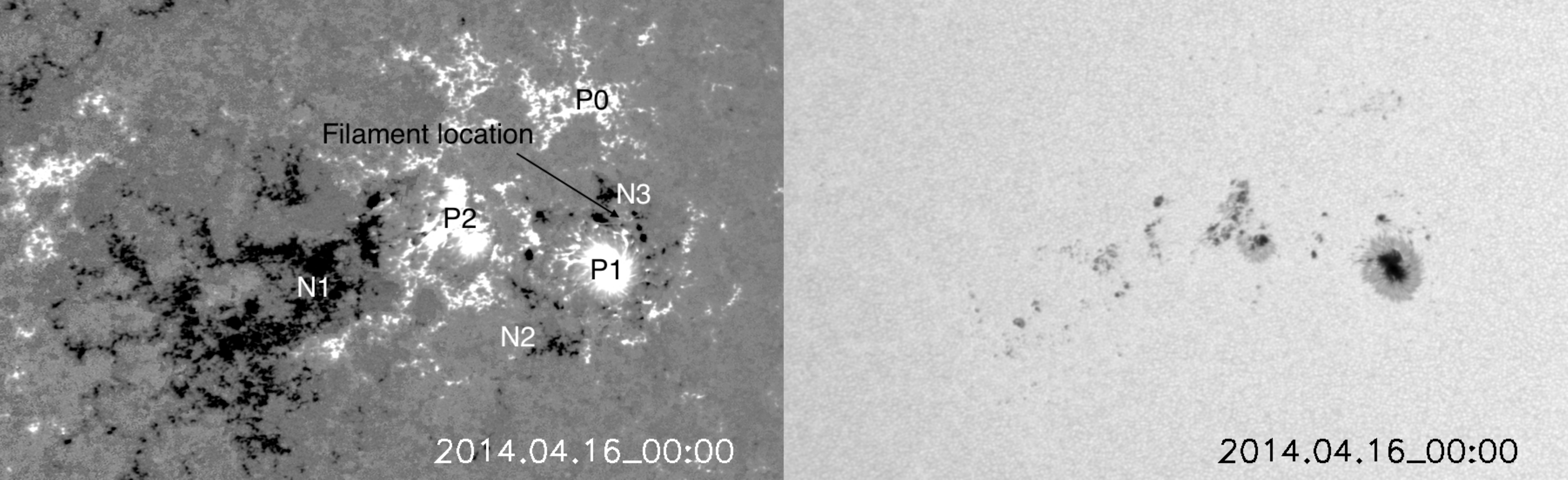}
\caption{ HMI SHARP data projected and remapped to a cylindrical equal area (CEA) Cartesian coordinate system.  Distribution of the radial component of the magnetic field (left) and of the HMI continuum (right). The FOV is [901 $\times$ 551] CEA pixels. The color scale for the magnetic field is saturated at $\pm$500 Gauss and black/white indicate negative/positive magnetic field. P0, P1, and P2 (resp. N1, N2, and N3) indicate different relevant positive (resp. negative) magnetic field distributions discussed in the text. The temporal evolution is shown in  Movie~3 available online. }
\label{Fig:Br-continuum}
\end{center}
\end{figure*}

%--------------------------------------------------------------------
\section{Observations}

The active region NOAA 12035 appeared at the East limb on  2014 April 11 with a $\beta$ magnetic configuration and crossed the West limb on 2014 April 23. During its disk passage it produced many small--to--medium class solar flares. 
The active region turned into a $\beta\gamma$ magnetic configuration on 2014 April 13.
During the disk passage on 2014 April 15 and 16, the active region (located S15, E20 to E08) produced  eruptive and compact flares, respectively.
The description of these confined and eruptive flares is given in Table 1. 

The flares on April 15 and 16  were  observed by SDO with a high spatial and temporal resolution. 
In the current study, we  used data acquired by the {\em Atmospheric Imaging Assembly} \citep[AIA][]{AIA}  
on-board {\em Solar Dynamic Observatory} \citep[SDO][]{SDO}. AIA observes the full Sun with different 
filters in EUV and UV spectral lines with a cadence of 12 sec and a pixel size of $0\farcs 6$. For this 
study, we used  AIA 171~\AA\ and 304~\AA\ data.

For the chromospheric observations of  one event on 2014  April 16, we used the 15 cm Coud\'e telescope equipped with a  H$\alpha$ filter
from  Aryabhatta Research Institute of Observational Sciences (ARIES), Nainital, India. These images are acquired with a cadence of 10 seconds and have a pixel size of  1 arcsec. 

To study the magnetic topology of the active region, we used the data acquired by the Helioseismic 
and Magnetic Imager \citep[HMI,][]{HMI} aboard {\it SDO}. HMI measures the photospheric magnetic field
of the Sun with a cadence of 45 sec and a  pixel size of $0\farcs 5$. 

Finally to search for the possible CME associated with the flares, we used the LASCO  C2 coronagraph data aboard 
the {\em }SOHO mission \citep{Bru1995}.

\subsection{LASCO/CME Observations}

All the eruptive flares occurred on April 15. The CMEs associated with two of these eruptive flares are presented in Figures~\ref{Fig:OBS-CMEa} and \ref{Fig:OBS-CMEb}. The CME associated with the C8.6 X-ray flare at 09:15~UT  is first seen in LASCO C2 coronagraph at 10:36 UT (Figure~\ref{Fig:OBS-CMEa}), and is characterized  by a narrow  angular width of 27 degrees and an average speed of 274 km s$^{-1}$.  The CME associated with the  C7.3  X-ray flare that occurred at 17:53 UT is visible in LASCO C2 field of view (FOV) at 20:00~UT (Figure~\ref{Fig:OBS-CMEb}), and  has an angular width of 179 degrees  and an average speed of 360 km s$^{-1}$. 

On April 16 no CMEs associated  with the flares in the active region are observed. One example of the corona observed two hours after the flare that occurred at 12:42~UT is presented  in  Figure~\ref{Fig:OBS-CMEc} to show that no CME is detectable. However, we note that two CMEs  are recorded on April 16  (see Table 1). After a detailed inspection of the LASCO movies we identified  a poor CME around 06:30~UT that it is  too early to correspond to the flare at 06:37~UT, and a narrow CME directly towards  the south at 20:00~UT that is  again  too early to correspond to the M1.0 flare. These CMEs could correspond to jet activity that characterize the eastern part of the active region.

\subsection{SDO/AIA Observations}

All the flares of the AR 12035 considered in this study and listed in Table~1 were well observed by  the AIA instrument on board  SDO. Apart from the last one (flare~26 in Table~1), they are all low energy events, and correspond,  for the strongest ones, to C--class flares. During 2014 April 15 and 16, the active region produced six eruptive and twenty confined flares. Two of the eruptive flares, productive of CMEs,  that occurred on April 15 are shown in the first two rows of Figure~\ref{Fig:OBS-15-04} (in AIA 304~\AA\ pass band).

Figure~\ref{Fig:OBS-15-04} presents the  environment of the  region around the time of the flare,  and the black box  in panel $b$ is focused on the AR 12035.   A zoom of its evolution is presented in Figure \ref{Fig:filament}. As an example, the image at 09:15~UT shows a dark north-south oriented filament that has been activated a few minutes before and, consequently bright arcades are observed around it. The flare emission reaches its maximum at 09:23~UT while at 09:37~UT a round shape brightening can also be observed (Figure~\ref{Fig:OBS-15-04b} and accompanying Movie~1). Finally, between  09:40~UT and 10:03~UT  dark strips  are seen to cross the active region from West to East (see Figure~\ref{Fig:OBS-15-04c} and accompanying Movie~1).

The second flare that we consider here occurred at 17:53~UT, when we see the activation of the filament that started to be more east-west aligned and with a second half-circle shaped filament at the north of it (Figure~\ref{Fig:OBS-15-04e}).  Movie~1 shows that, after a first failed eruption of the southern threads of the filament, the main body of the filament starts to erupt at 17:51~UT, when  circular  bright arcades on the west of the filament are also visible (Figure~\ref{Fig:OBS-15-04e} and Movie~1). During the eruption the  filament  interacted with the environment and, similarly to the other eruptive flares, resulted in the ejection of plasma, visible as dark stripes around 18:33 UT (Figure~\ref{Fig:OBS-15-04e} and Movie~1).  These dark, filamentary eruptive structures that can be clearly seen in Figures~\ref{Fig:OBS-15-04c} and \ref{Fig:OBS-15-04f} had a duration of about 45 minutes, and eventually produced the CMEs.
%One of the interesting phenomena observed in the case of these four eruptive flares is the presence of long dark  line-shaped structures of cool material crossing the region from West to East  after the flare onset.

During the two days of observation the filament(s) evolved from being constituted by two separated filaments on April 15 ---one relative-straight and north-south at 09:15 UT oriented  and an upside down U  at the north of the first one well visible at 17:00 UT --- to a complete east-west oriented  sigmoidal  filament  on April 16  at 05:40 UT  (see Figure \ref{Fig:filament} and accompanying Movie~2). All failed eruptions observed on April 16 were initiated by an  asymmetric failed eruption of the southwestern part of the sigmoidal filament. 

One example of a compact flare  that occurred on April 16 is shown in the bottom row of Figure~\ref{Fig:OBS-15-04}.  Around the time of the onset of the flare, i.e., at 12:42~UT, we observe an oval shape of brightening  around the AR 12035 with inside the dark sigmoid and many filamentary structures in its southwestern  end (Figure~\ref{Fig:OBS-15-04h} and  also Figure \ref{Fig:filament}). Until this time the dynamics is similar to what was observed the day before. However,  at 13:00~UT a bright overlying arcade is seen  over the AR 12035  and the dark material inside stops to rise (Figure~\ref{Fig:OBS-15-04i} and Movie~2).
The eruption concerned only the southern part of the  sigmoid and did not succeed to drive all the sigmoid to erupt. The two other failed eruptions,  at 10:42~UT and at 20:00~UT, followed  the same scenario. These three events lasted 15 minutes each. The eruption  of 10:42~UT is well observed in H$\alpha$ and is discussed in detail in the next Section.

As a final remark, we note that recurrent jet activity is recorded at the southeast of the flaring activity.  The study of this jet activity is outside the scope of the present paper that focuses on the transition from eruptive to confined flares in AR 12035.

\subsection{H$\alpha$ Observations}

In this subsection we discuss the failed eruption that occurred on April 16 at 10:42~UT and that is well observed in H$\alpha$ from ARIES, Nainital.

The H$\alpha$ image taken at 10:34~UT on April 16 (Figure~\ref{Fig:Halpha}), shows the S-shaped filament  (S) in the north of the active region, which was formed between April 15 and April 16 (see Figure~\ref{Fig:filament}).  
Around 10:38 UT, the filament started to  be activated, and at around 10:46~UT it broke in its center. The northern part of the filament remained in its original condition, while the broken part of  it consisted of  many threads (T)  that are visible in H$\alpha$ at 10:48~UT  (Figure~\ref{Fig:Halpha}), when the filament started to erupt in the west direction. 
However, the broken filament's southern foot point remained fixed. Eventually, the erupted part of the filament fell back on the solar surface, resulting  in a failed eruption. Together with the filament eruption close to the breaking location of the filament, we observe the maximum flare brightening  at 10:51~UT. Later on, we observe  two flare ribbons  (R1, R2)  at 10:53~UT. Finally, we note that the dark H$\alpha$ structure with a fan-like shape  (F) did not expand after 10:58~UT (see Figure~\ref{Fig:Halpha}).

The AIA 171~\AA\ observations confirm the failed eruption (Figure~\ref{Fig:Halpha}, bottom row).  The filament is visible  in absorption  with a S-shaped  at 10:47~UT and with a side-view of the arcade overlying the western part of the filament during the eruption at 10:54~UT. The ribbons appear as  bright structures along the foot points of the arcades.

\begin{figure*}
\begin{center}
\subfigure[]{
\includegraphics[width=.48\textwidth,viewport= 50 173 530 520,clip]{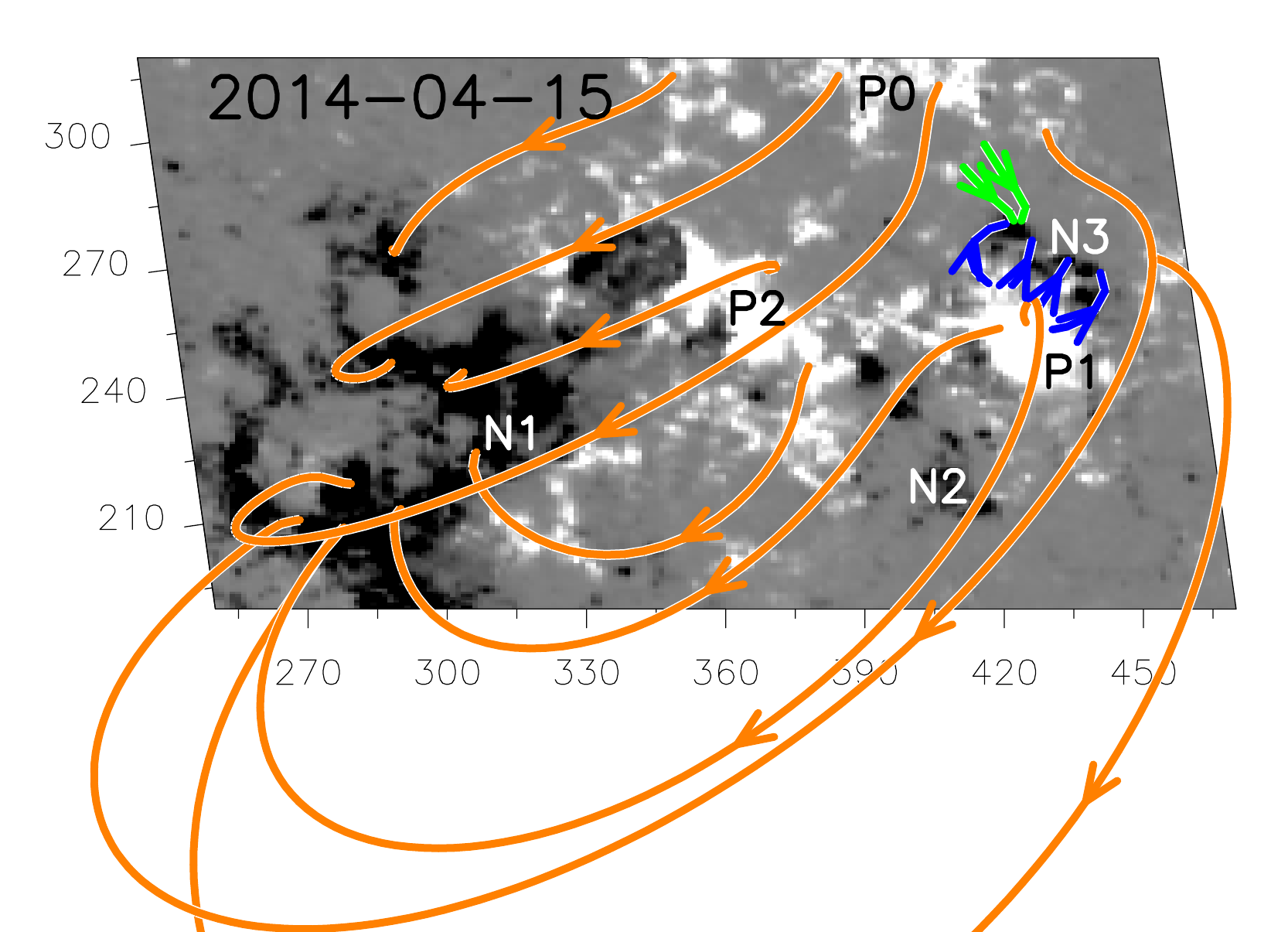}\label{Fig:T-Overview1}}
\subfigure[]{
\includegraphics[width=.48\textwidth,viewport= 50 200 530 520,clip]{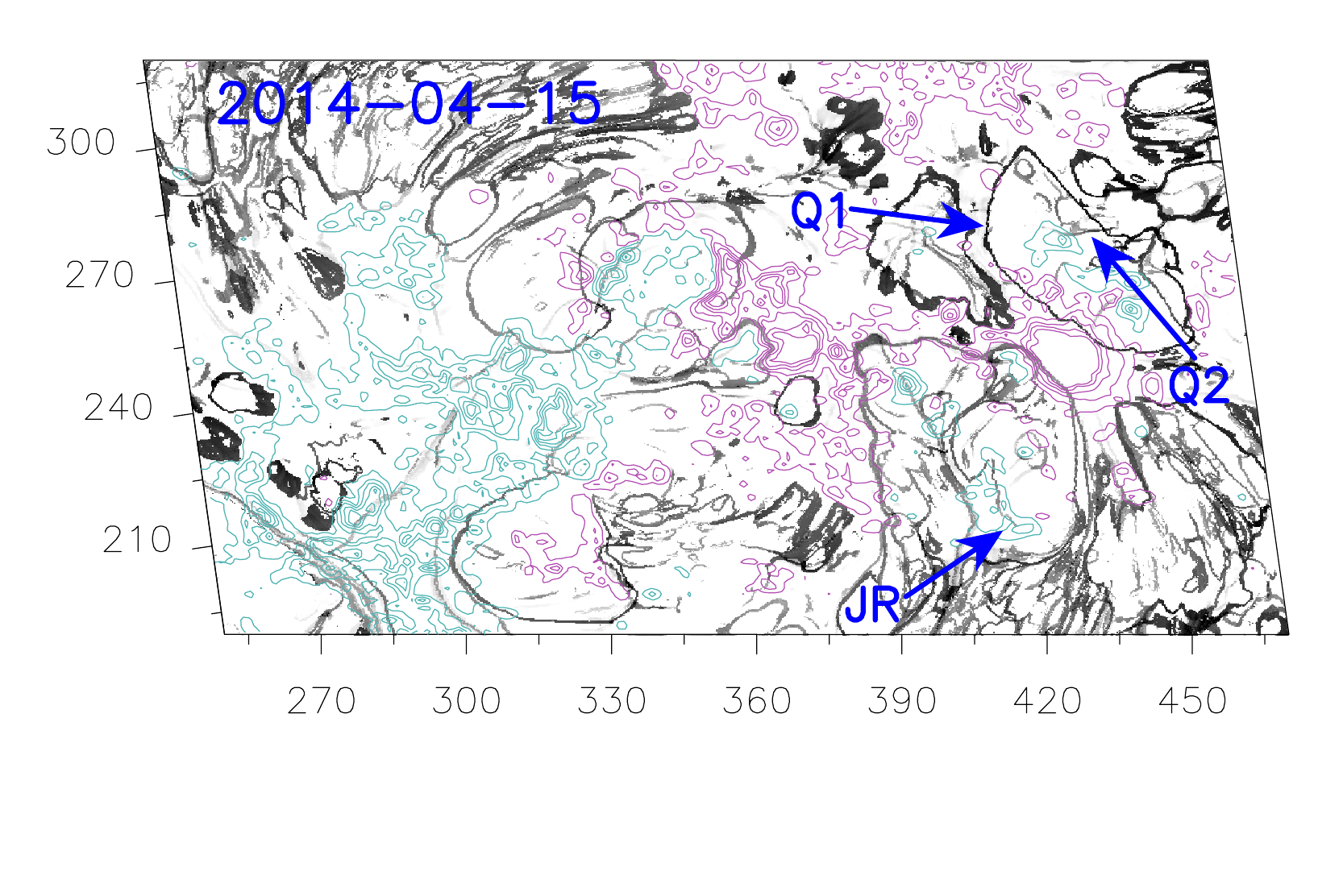}\label{Fig:T-Overview2}}
\subfigure[]{
\includegraphics[width=.48\textwidth,viewport= 50 173 530 520,clip]{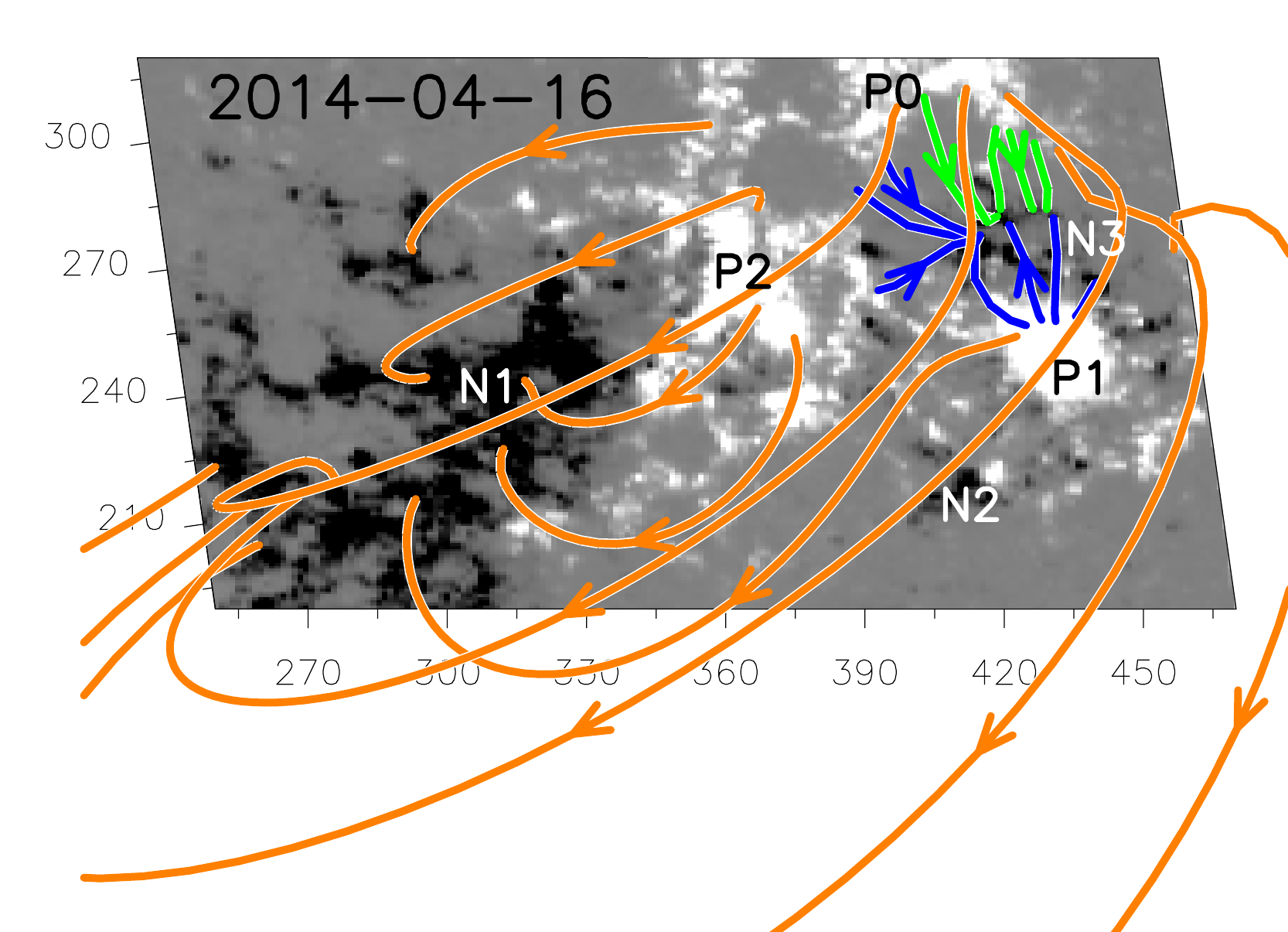}\label{Fig:T-Overview3}}
\subfigure[]{
\includegraphics[width=.48\textwidth,viewport= 50 200 530 520,clip]{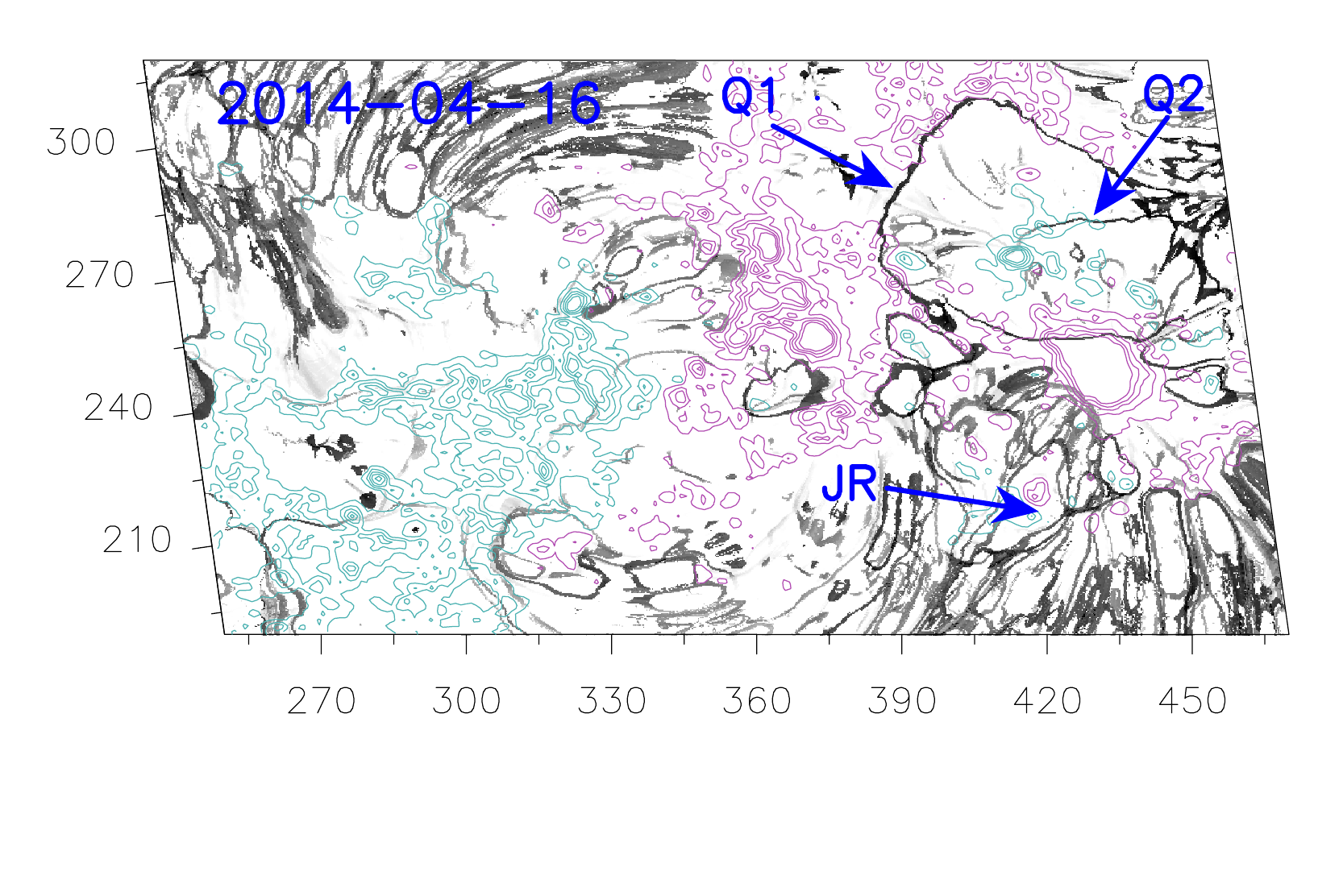}\label{Fig:T-Overview4}}
\caption{Magnetic field distribution (line-of-sight component) of AR 12035 together with  some representative potential field lines (left panels) and QSL maps  (right panels) for April 15 (top) and April 16 (bottom). The QSL maps (gray color scale) are computed at $z=0.4$~Mm above the photosphere (see Section~\ref{Sect:QSL}). The color scale for the magnetic field  is saturated at $\pm$300 Gauss and black/white indicate negative/positive magnetic field, while the magenta/cyan contours indicate positive/negative magnetic field of $\pm$100, $\pm$300, $\pm$500, $\pm$700 and $\pm$900 Gauss. The axis indicate Mm from the bottom-left corner of the larger remapped HMI/LOS magentogram used to perform the potential field extrapolation (see Section~\ref{Sect:MFExtr} for more detail). Q1 and Q2 indicate the QSLs relevant for the eruption, while JR indicate the region where the recurrent jets occurred (see Section~\ref{Sect:QSL}). }
\label{Fig:T-Overview}
\end{center}
\end{figure*}

\begin{figure*}
\begin{center}
\subfigure[]{
\includegraphics[width=.32\textwidth,viewport= 45 180 538 590,clip]{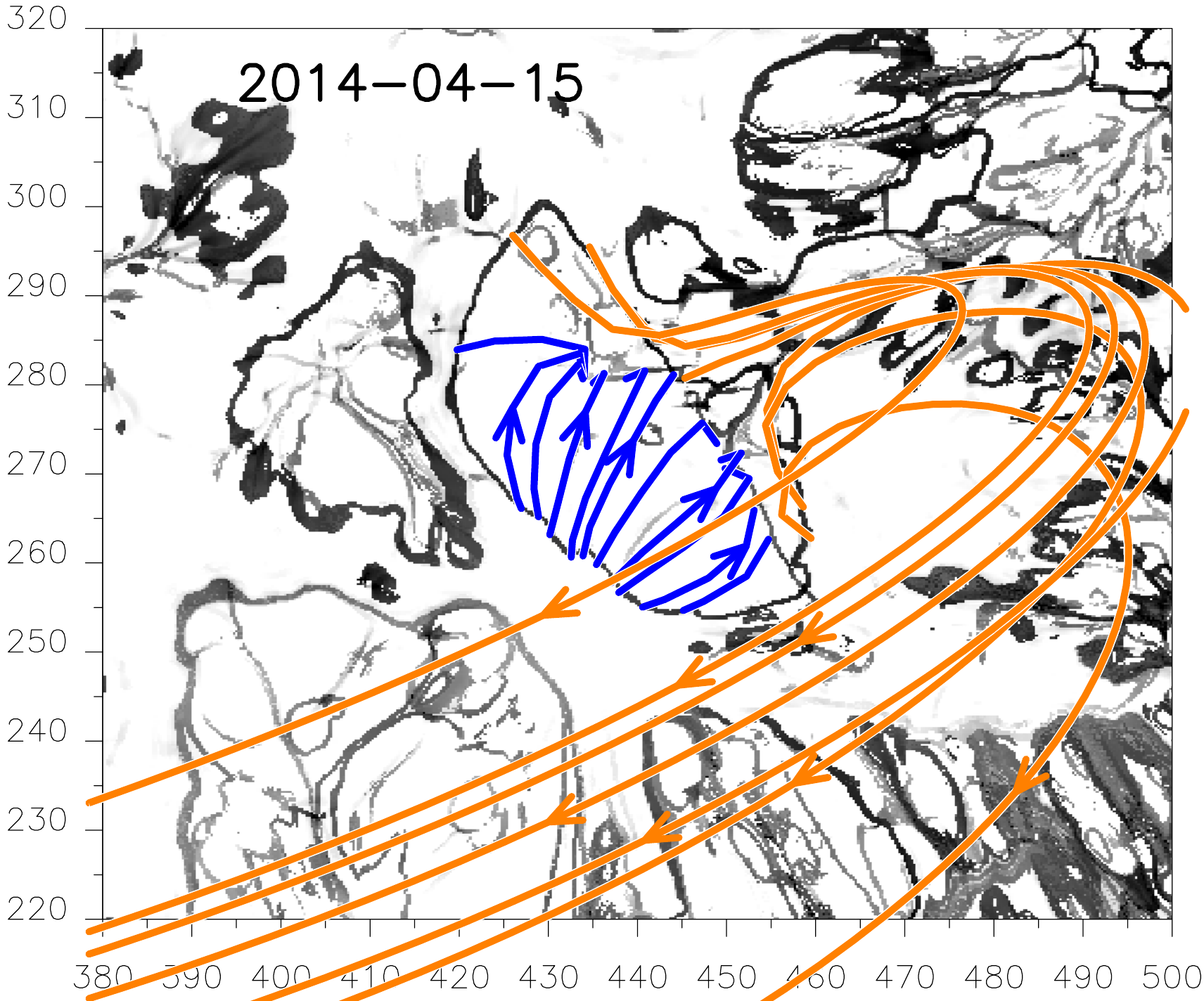}\label{Fig:T-QSL1}}
\subfigure[]{
\includegraphics[width=.32\textwidth,viewport= 45 180 538 590,clip]{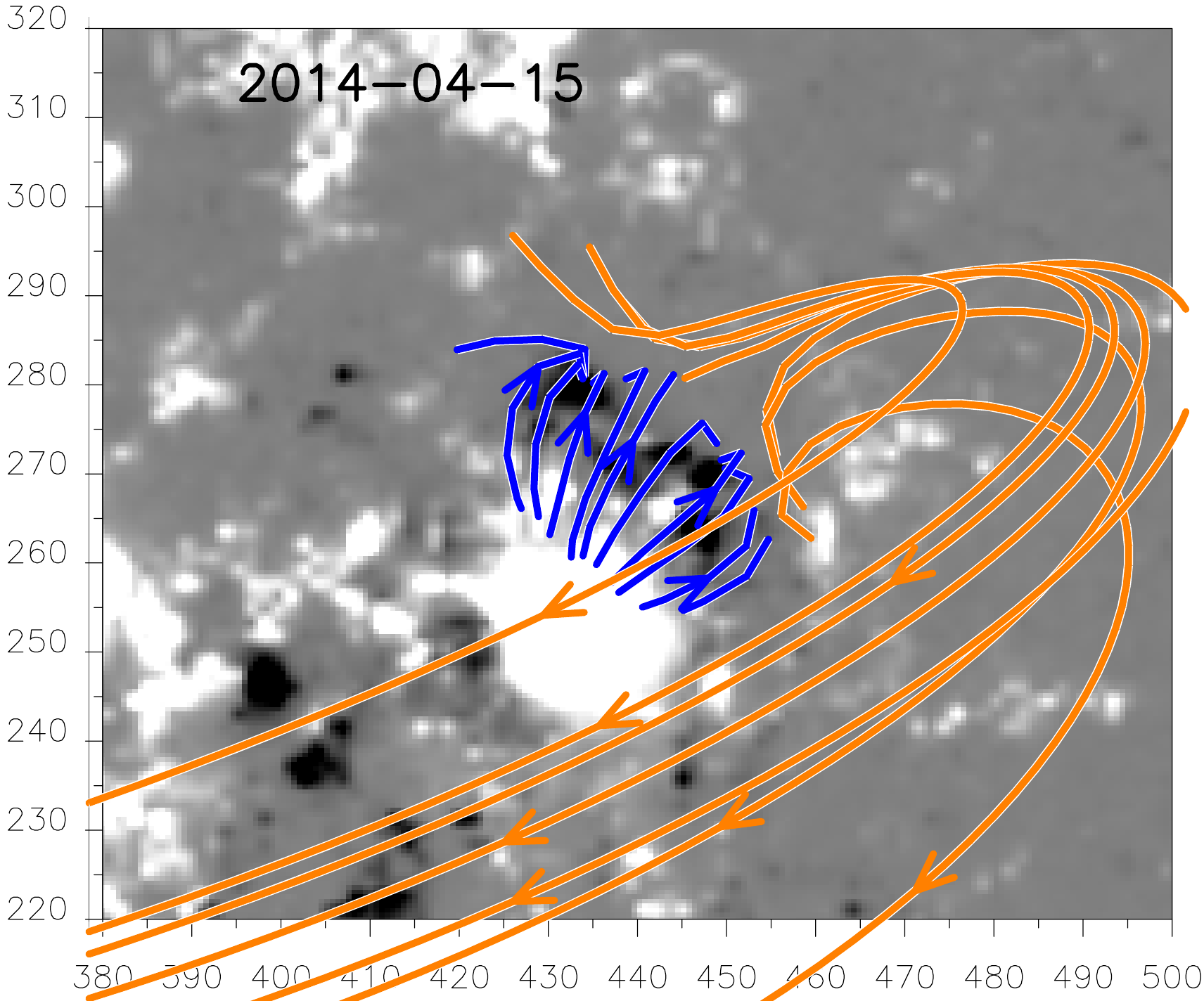}\label{Fig:T-QSL2}}
\subfigure[]{
\includegraphics[width=.28\textwidth,viewport= 70 190 495 610,clip]{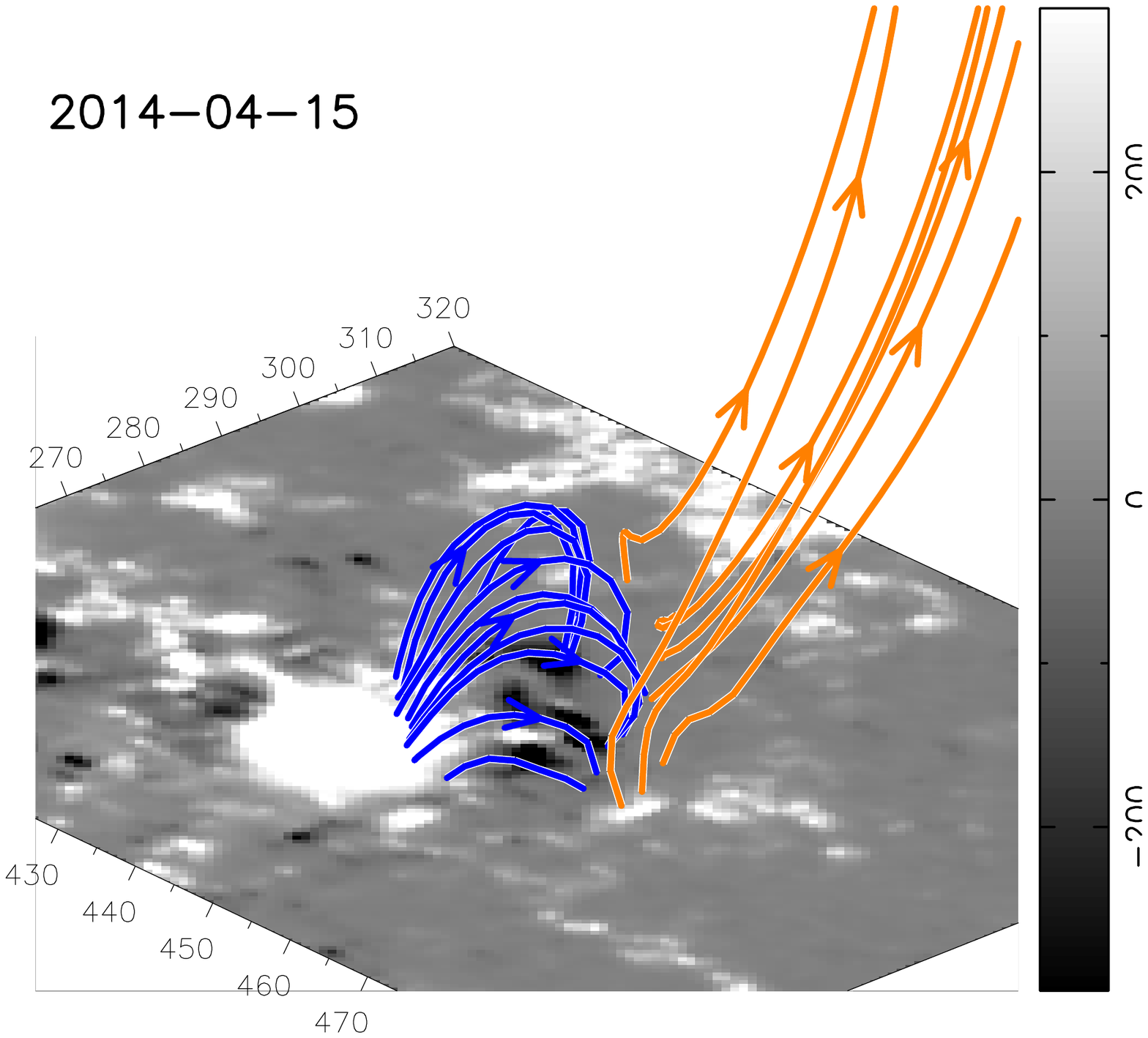}\label{Fig:T-QSL3}}
\subfigure[]{
\includegraphics[width=.32\textwidth,viewport= 45 180 538 590,clip]{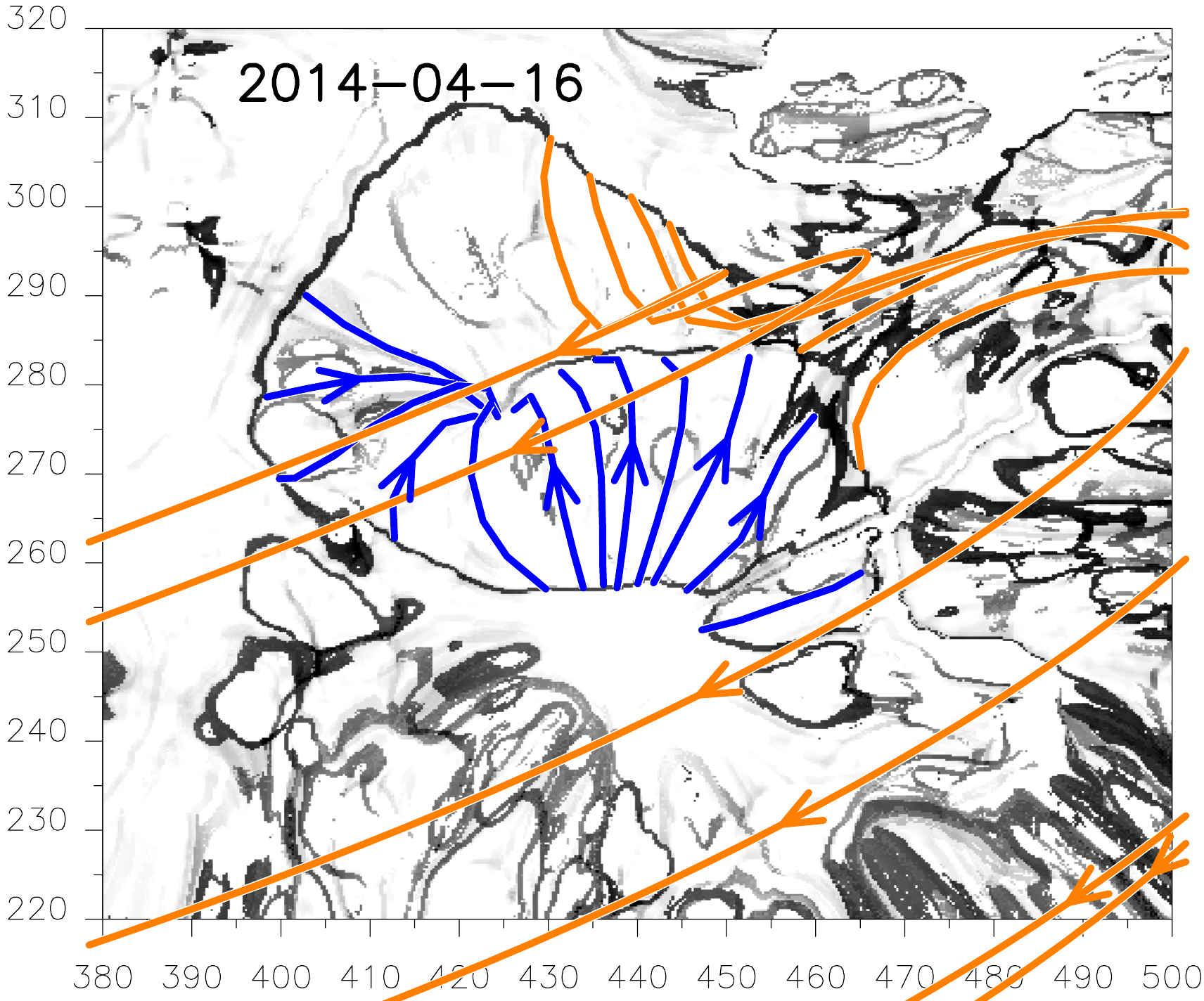}\label{Fig:T-QSL4}}
\subfigure[]{
\includegraphics[width=.32\textwidth,viewport= 45 180 538 590,clip]{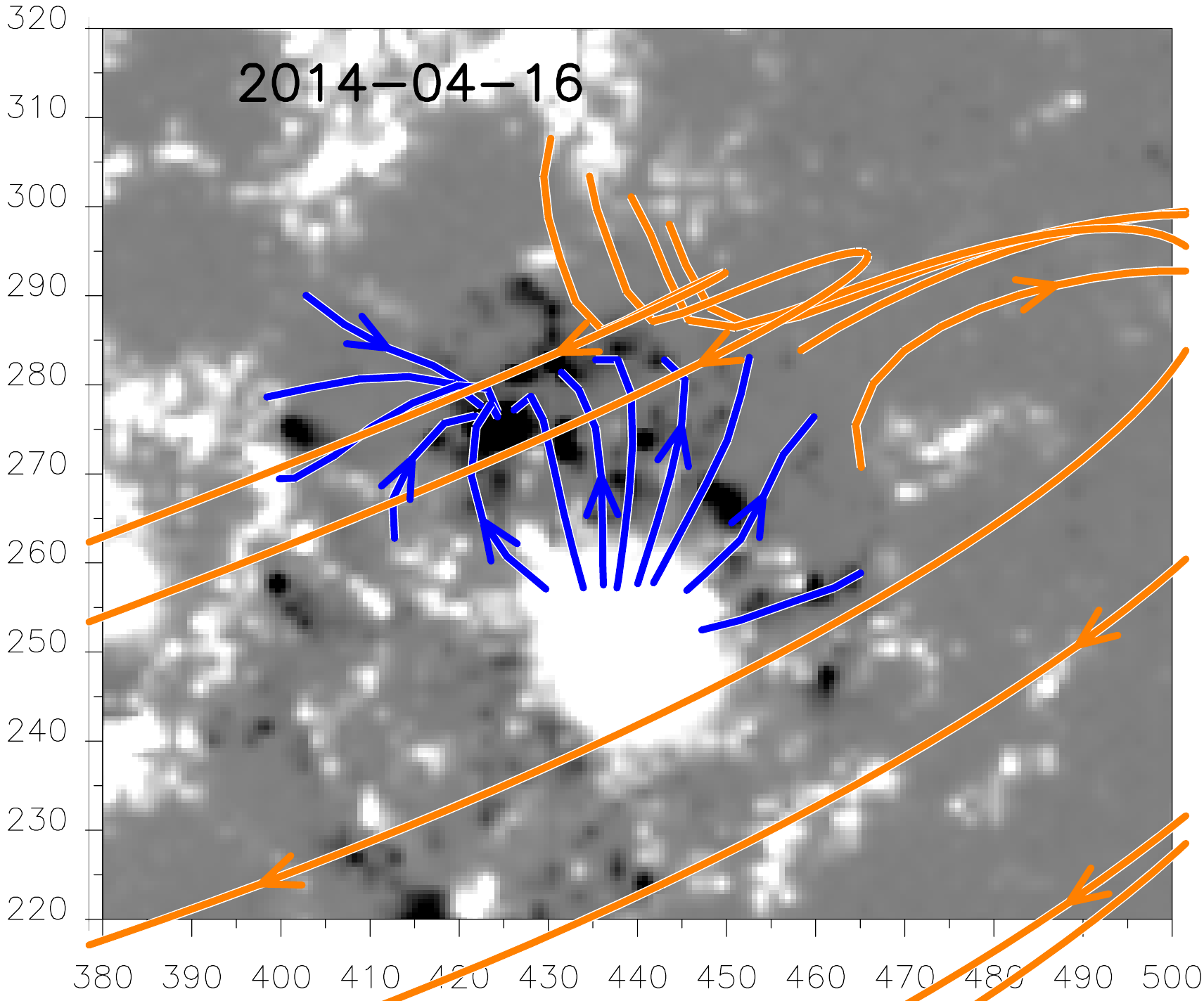}\label{Fig:T-QSL5}}
\subfigure[]{
\includegraphics[width=.28\textwidth,viewport= 70 190 495 610,clip]{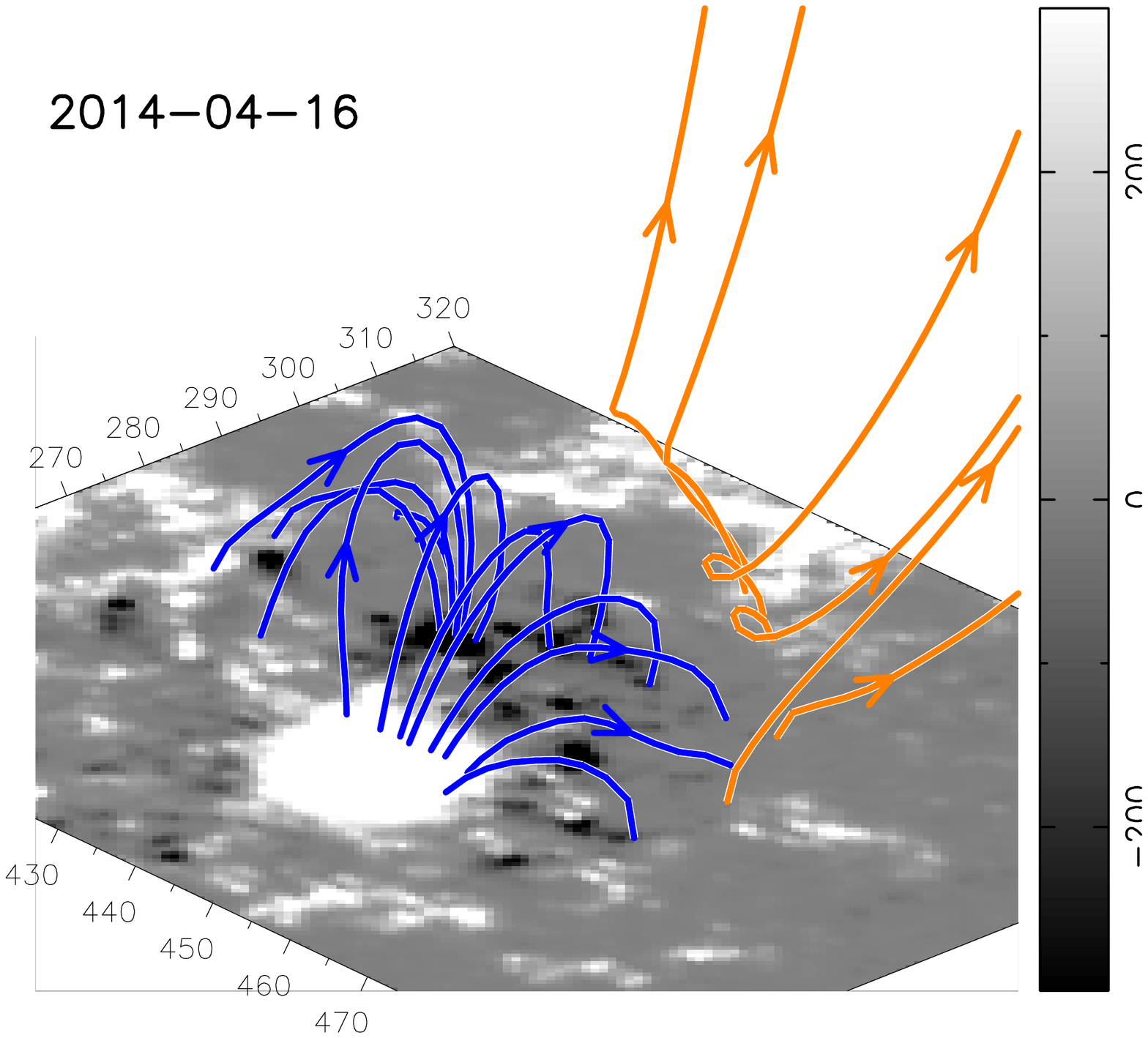}\label{Fig:T-QSL6}}
\caption{Zoomed-in views of the distribution of  the QSL maps at $z=0.4$ Mm (gray color scale, left) and of the photospheric (line-of-sight component) magnetic field (gray color scale, middle, right) as well as selected magnetic field lines for April 15 (top) and  April 16 (bottom).  For better visibility field lines in panels (c,f) are vertically stretched by a factor 2. The color scale for the magnetic field  is saturated at $\pm$300 Gauss and black/white indicate negative/positive magnetic field.}

\label{Fig:T-QSL}
\end{center}
\end{figure*}

\subsection{Evolution of the magnetic field}
\label{Sect:MF}

On April 15, the active region had an overall bipolar structure characterized by a positive leading polarity P1 and a following negative one N1 (Figure~\ref{Fig:Br-continuum} and accompanying Movie~3, left panel). The leading polarity appeared to be constituted by a preceding compact flux distribution P1, coinciding with the umbra of the leading sunspot (Figure~\ref{Fig:Br-continuum}, right panel), followed by a more disperse polarity P2.  The positive polarity P1 is surrounded by a moat region with frequent bipole flux emergence and cancellation. Consequently, the positive polarity P1 is surrounded by two negative flux distributions, indicated as N2 and N3 in Figure~\ref{Fig:Br-continuum}, left panel.  
The continuum intensity (Movie~3, right panel) shows that the sunspot P1 is rotating in the clockwise direction by an angle of about 35 degrees during the two days of observations (and about 80 degrees between April 14 and 18, not shown in the Movie). 

Between April 15 and 16 we observe the emergence of new magnetic flux between the  dispersed positive flux P2  and the following negative polarity N1 (see Movie~3). This region corresponds to the area of an arch filament system (AFS) visible in Figure~\ref{Fig:OBS-15-04e}. As a result of this process, part of the positive dispersed flux that constitutes the leading polarity is annihilated  and the separation between the negative N1 and positive P2 flux distribution increases (see Movie~3). Furthermore, the leading polarity is now characterized  by two compact distributions of positive flux that are well separated 
%between them 
(see P1 and P2 in Movie~3 and Figure~\ref{Fig:T-Overview3}). 

Starting from about 19:00~UT on April 15  a succession of bipoles with a larger negative polarity and a weaker positive one is seen to emerge in the north of P1 leading to an accumulation of flux in N3 (see Movie~3). Subsequently, we observe a northeast migration  with a counter clockwise rotation of the newly emerged flux N3. Therefore, there is a strong shear between the clockwise rotating polarity P1 and the counterclockwise rotation of N3.

Contemporaneously to this migration, small concentrations of magnetic flux are seen to spread from the compact leading polarity P1 in all directions. As a result part of the flux of P1 is canceled with the negative fluxes N3 and N2. The recurrent jets  (visible in  Movie~1) probably originate from the cancellation of N2 and P1 that may lead to magnetic reconnection producing the observed jets around the location of N2.

By 10:24~UT on April 16  (Movie~3 and Figure~\ref{Fig:T-Overview3}) the positive polarity of the AR is constituted of three separate (more or less compact)  distributions of positive flux (see P1, P2, P3 in Figure~\ref{Fig:T-Overview3}) with a negative intrusion N3 at the north of the leading compact one. 

The filament that is the subject of this study is located along the PIL between the compact positive polarity P1 and the negative flux distribution N3  (Arrow in Figure~\ref{Fig:Br-continuum}, left panel and blue arcades in Figure~\ref{Fig:T-Overview}). 

\section{Topology of the magnetic field }
\label{Sect:MFT}

In this Section we describe the key topological structures of the active region between April 15 and 16, i.e., between the time period when the nature of the flares changed from eruptive to confined.

\subsection{Potential magnetic field extrapolations}
\label{Sect:MFExtr}

To study the connectivity of the different flux domains and their evolution we computed  a potential magnetic field model of the AR (Figure~\ref{Fig:T-Overview}, left columns). Potential configurations give robust information on the topological structures of the coronal field such as separatrices and quasi-separatrices \citep[see Section~\ref{Sect:QSL} and][]{Dem1996}

Since  we are mainly focused on the connectivity of the active region,  we perform the potential extrapolation using a larger FOV provided by the HMI LOS-magnetograms that includes the neighboring active regions rather then the much smaller FOV provided by the HMI SHARP data product. To this purpose the HMI LOS-magnetograms of AR 12035 (and its neighboring active regions) taken at 10:24~UT on April 15 and April 16 have been re-mapped to the disk-center using the mapping software available  through \textit{SolarSoft}. As a result of this process the AR 12035 is rotated so that its center is located along the central meridian. 
During this process we also decreased the resolution of the images from the $\sim 0.5$ arcsec of HMI to $\sim 2$ arcsec. The subregion of the de-rotated magnetogram (containing both AR 12035 and the neighboring active regions) is then  inserted at the center of a 8 times larger grid padded with zeros. The potential field extrapolation is  performed by applying  the fast Fourier transform method of \cite{Ali1981} on this larger grid. 

As Figure~\ref{Fig:T-Overview} (left panels) shows the large scale magnetic field is indeed bipolar as discussed in Section~\ref{Sect:MF}, but the part of the active region that displays an increased level of activity is characterized by a more complex connectivity. Essentially, four flux domains are observed: the first connecting the north-most part of the positive polarity P0 to the negative flux N3 at its south (green field lines in Figure~\ref{Fig:T-Overview1}),  the second connecting the negative polarity N3 with the leading compact positive polarity P1 (blue field lines), the third connecting this latter with the negative flux N2 at the southeast of it (connecting field lines not shown), and the last one is the large scale field that connects the positive polarities P0, P1 and P2 to the following negative one (N1, orange field lines). 

The anemone-like structure (blue-green field lines) is embedded in a bipolar field resulting in a \textit{breakout}-like magnetic field configuration, and  evolves from an northeast-southwest elongated structure on April 15 to a more circular one on April 16 (see Figure~\ref{Fig:T-Overview}, left panels).

\begin{figure*}
\begin{center}
\subfigure[]{
\includegraphics[width=.48\textwidth,viewport= 400 115 769 375,clip]{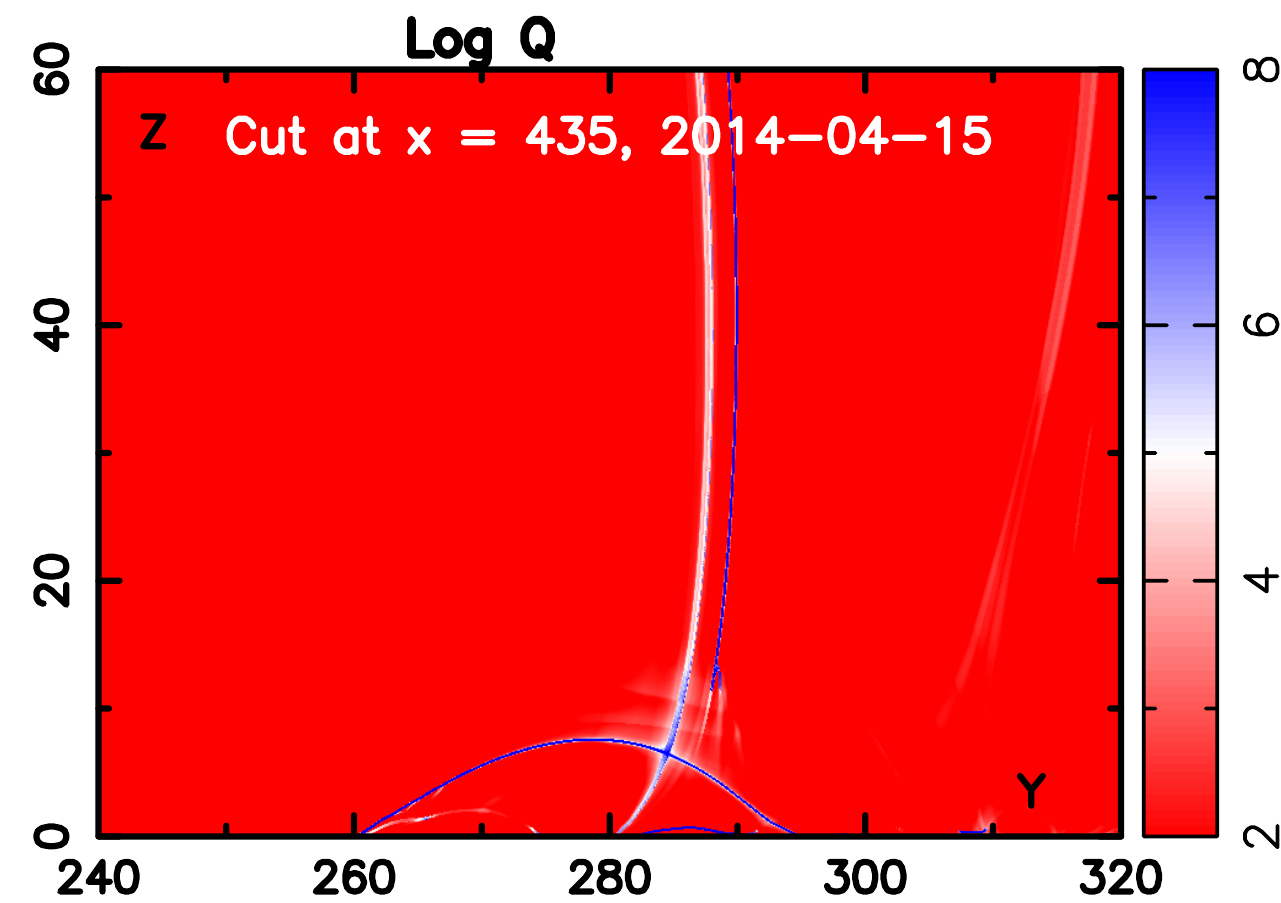}}
\subfigure[]{
\includegraphics[width=.46\textwidth,viewport= 10 73 408 362,clip]{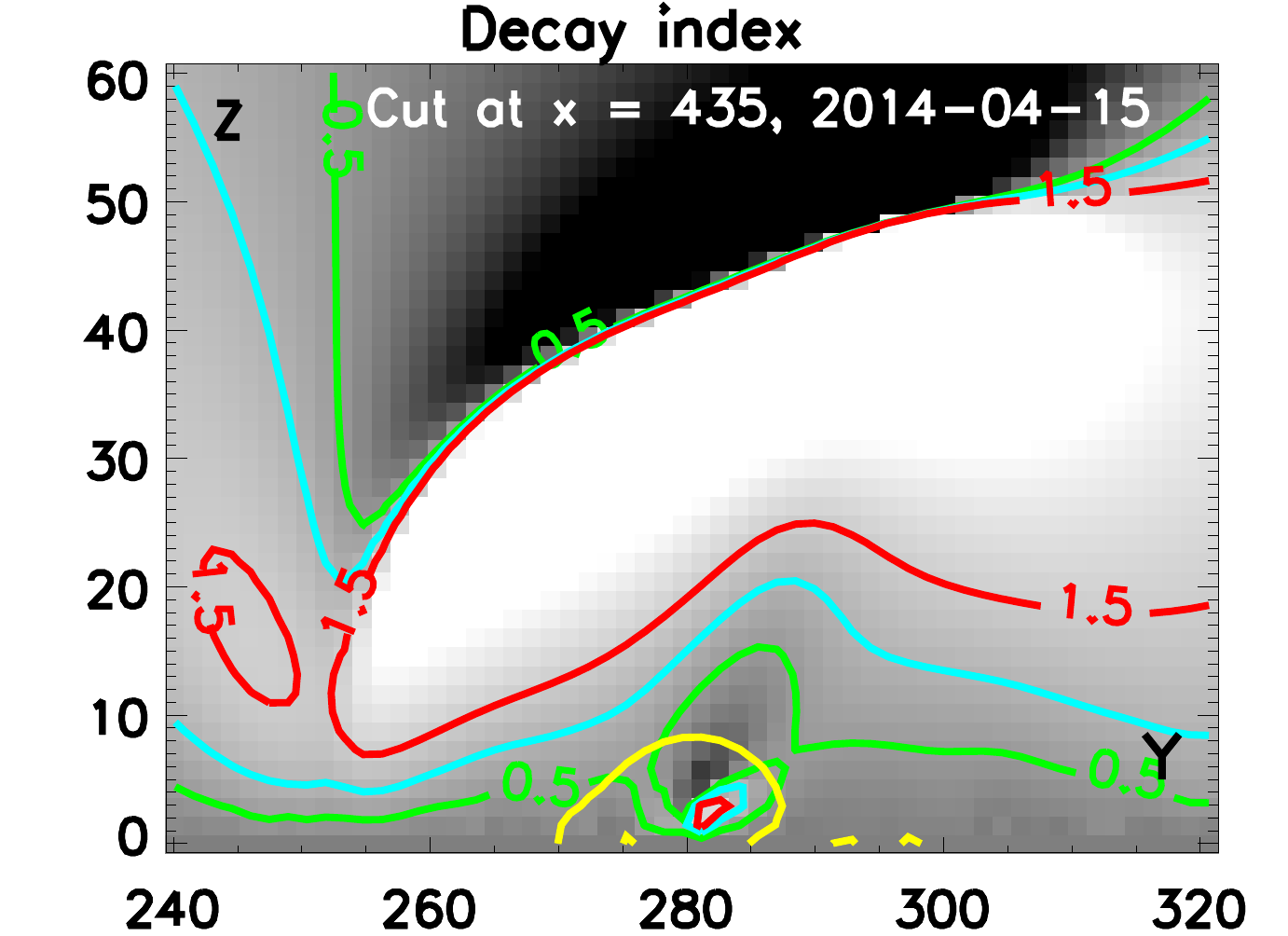}}
\subfigure[]{
\includegraphics[width=.48\textwidth,viewport= 400 115 769 375,clip]{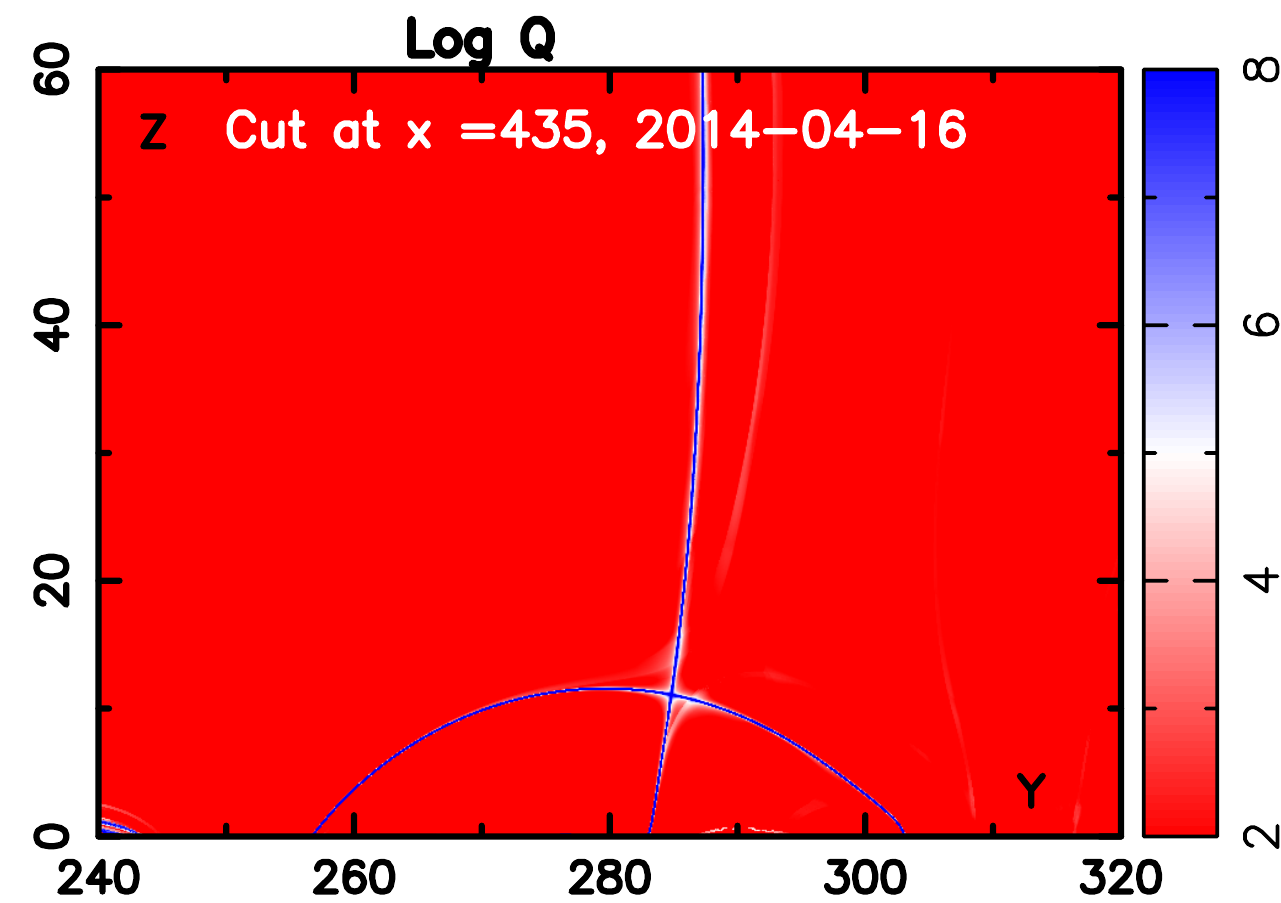}}
\subfigure[]{
\includegraphics[width=.46\textwidth,viewport= 10 73 408 362,clip]{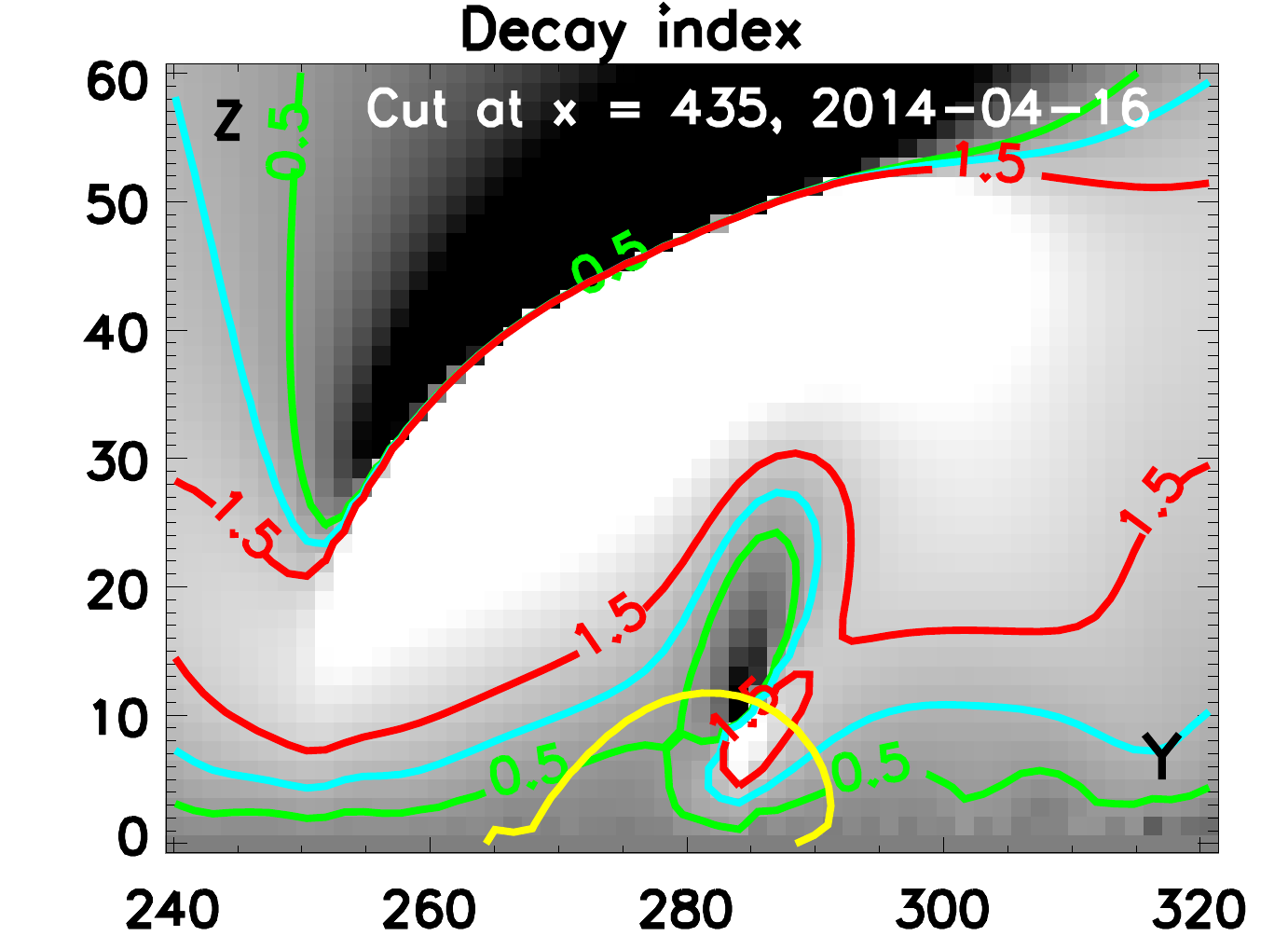}}
\caption{Projected 2D view of the squashing degree \textit{Q} (left panels, see Section~\ref{Sect:MFT}) and of the decay index (gray color scale, right panels) along a plane passing through x = 435 (see Figure~\ref{Fig:T-QSL}).  The HFT corresponds to the intersection of the two high $Q$ regions in the left panels. The green, cyan and red contours on the right panels indicate isocontours of decay index $n= 0.5, 1,$ and $1.5$, respectively. The yellow contour indicates the polarity inversion line. Axis units are in Mm from the bottom-left corner of the larger remapped HMI/LOS magnetogram used to perform the potential field extrapolation (resp. from the photosphere) for the abscissa (resp. the ordinate).  }
\label{Fig:T-QSL_vert}
\end{center}
\end{figure*}

\subsection{Quasi-separatrix layers}
\label{Sect:QSL}

Quasi-separatrix layers \citep[QSLs,][]{Dem1996} are thin layers characterized by a finite, but sharp, gradient in the connectivity of the magnetic field, and are  defined as regions where the squashing degree $Q$  is \textit{large}  \citep{Tit2002}. QSLs are also locations where current layers easily develop, where (slip-running) magnetic reconnection can occur \citep{Aul2006, Jan2013, Dud2014a}, and they often coincide with the position of the flare ribbons \citep{Sav2012,Sav2015,Sav2016, Zhao2014,Zhao2016}. 

In this work we compute the $Q$-factor using the latest version of the topology tracing code \citep[topotr,][]{Dem1996}, where the formula of \cite{Par2012} is implemented. To this purpose we define the plane at $z=0.4~$Mm as the seed plane from which the field lines  are traced. 

On April 15 an elongated  fan-type QSL (Arrow~Q1, Figure~\ref{Fig:T-Overview2}) surrounds the negative magnetic field distribution N3 at the northwest of the compact leading positive polarity P1 and embeds the portion of the PIL where the filament is located. This QSL  essentially encloses and separates the anemone-like structure (blue-green field lines) from the global/large-scale field (orange field lines) of the active region.  A spine-like QSL that starts from the northwest part of the fan-QSL and intrudes towards the central part of it is also observed (Arrow~Q2, Figure~\ref{Fig:T-Overview2}). Field lines that originate at the north of the spine-QSL connect to the north-most positive polarity P0  (green lines), while the ones that originate at the south of it  connect to the compact leading positive polarity P1 (blue lines, Figures~\ref{Fig:T-QSL1} and \ref{Fig:T-QSL2}). These latter are the ones that enclose the filament that is the object of this study. 

On April 16 the fan-QSL (Arrow~Q1, Figure~\ref{Fig:T-Overview4}) displays a more circular and less elongated shape, while  the spine-QSL (Arrow~Q2, Figure~\ref{Fig:T-Overview4}) now originates from the center of the fan-QSL circle and extends westward. This magnetic field configuration indicates the presence of an elongated, locally two-dimensional, hyperbolic flux tube (HFT) that separates the two lobes of the anemone-like magnetic field configuration from each other and from the overlying field. This is confirmed by the vertical distribution of $Q$ along the plane passing through the spine-QSL (Figure~\ref{Fig:T-QSL_vert}, left panels). The 2D cuts show that a local `null-point like' configuration is achieved  in both cases, but the two lobes are more symmetric on  April 16 than they are on April 15. 

A second, less-pronounced, more-complex QSLs system is present around the region (Arrow~JR, Figure~\ref{Fig:T-Overview}) where the recurrent jets are observed. However, this latter does not intersect the QSL labeled as Q1  suggesting that the jet-producing region and the flaring region are not directly connected to each other (although propagating Alfv\'{e}n waves may still induce a causality connection between the two parts of the active region).

While QSLs are robust topological features essentially determined by the connectivity of the magnetic field their exact morphology depends on the actual magnetic field model used to compute them \citep{Sun2013}. To compute the $Q$-factor we used the simplest magnetic field compatible with the given boundary, i.e., the current-free magnetic field. Despite this very simple assumption we note (1) that the computed fan-QSL of the flaring region matches  well the circular flare ribbons,  (2) that similarly to the computed fan-QSL the flare ribbon actually crosses the compact leading polarity, (3) that a brightening is observed approximately at the location of the spine-QSL, and (4) that the jet-associated brightening do not cross the flare-associated ribbons (Movie~1).  This evidence suggests  that the  magnetic field model used is sufficient to capture the key features of the event. 

As a final remark, we note that the discrepancy between the computed fan-QSL (Figure~\ref{Fig:T-Overview}) and the circular brightening ribbon (Figure~\ref{Fig:OBS-15-04}) is probably also due to the simplistic magnetic field model used. This can be seen from Figure~10 of \cite{Sun2013} where the QSLs computed using both a potential field  and a non-linear force-free field (NLFFF) are compared. The fan QSL is relatively round in the potential field model, but displays a more sigmoidal shape in the NLFFF model that actually accounts for the shear present in the configuration. This is compatible with our configuration where the counterclockwise motion of the polarity N3 and the clockwise rotation of the sunspot P1 definitely introduced a degree of shear that the potential field model does not capture.  

\section{Discussion}

%\subsection{Decay index and torus instability}

A parameter that allows the estimation of the stability of a given magnetic field configuration is the decay index. Briefly, a magnetic flux rope embedded in an external magnetic field ($B_{\text{ex}}$) is unstable to perturbations if the axis of the flux rope has an height $z$ above the photosphere where the decay index of the external magnetic field:
\begin{equation}
n=-\frac{d \ln B_{\text{ex}}}{d \ln z},
\label{Eq1}
\end{equation}
is larger than a critical value $n_{\text{cr}}$, that depends on the morphology of the flux rope \citep{Dem2010, Zuc2015}, and is in the range $n_{\text{cr}} \simeq 1.1-1.75$ (see Introduction). %However, when the height of the filament and not of the axis of the flux rope is considered in the stability analysis then the critical value is closer to the lower end of the interval rather then to the upper end \citep{Fil2001, Zuc2014, Zuc2016}.   

To evaluate the stability of the magnetic field configuration we computed the decay index (using only the tangential component of the computed potential magnetic field) in all the volume above the flaring region. A vertical cut of decay index  along a plane passing through the approximate position of the HFT is shown in Figure~\ref{Fig:T-QSL_vert} (right panels).  The first conclusion that can be drawn from the Figure is that in the proximity of the HFT the decay index changes sign becoming negative  (and reaching very large, negative values) as already shown by \cite{Tor2007} and as expected from its definition (Equation~\ref{Eq1}). 

A second conclusion is that the large scale stability of the magnetic field (away from the HFT) does not change significantly between April 15 and 16. This can be deduced by comparing the height at which the decay index is larger than the ``nominal'' $n=1.5$ critical value. The decay index for both days shows an initial increase with altitude (i.e., the system is more prone to erupt), followed by a decrease at even larger altitudes (i.e., the system is torus stable). 

As previously discussed, both the circular-shaped photospheric-QSL (Q1, Figure~\ref{Fig:T-Overview}) and  the fan-spine-like distribution of $Q$ (2D vertical cuts of Figure~\ref{Fig:T-QSL_vert}) indicate the presence of a null-point topology in the corona. At the null-point the decay index (Equation~\ref{Eq1}) has a singularity and its validity is limited in this region. The distinction between torus instability or breakout-type reconnection as trigger mechanism for the eruption in this configuration is not at all straightforward  \citep{Kliem2014}.  Furthermore, in configurations with a vertical magnetic field, such as the one considered in this paper (see Figure~\ref{Fig:T-QSL}, right columns), the verticality of the field lines itself prevents any tension-related confinement even in a uniform field where the decay index is zero. Therefore, for this complex magnetic field topology the analysis of the decay index does not provide a useful criterion for eruptivity.  

For a configuration that displays a coronal null point, the breakout scenario is a valuable mechanism to trigger the eruption. In this scenario the eruption is triggered by the onset of magnetic reconnection, and the efficiency of it also depends on the mutual orientation of the reconnecting fields. \cite{Gal2007} performed a series of MHD simulations of a dynamical flux emergence experiment  aimed to study the role of the mutual orientation between the emerging flux rope and the overlying field. The authors have shown that when the two system are (nearly) anti-parallel substantial reconnection is observed, while this is not the case when the flux systems are (nearly) parallel. More recent simulations of dynamical emergence have shown that interaction of (nearly) anti-parallel flux systems leads to flux rope-like eruptions, while this is not the case for (nearly) parallel systems \citep{Arc2008, Arc2012, Lea2013,Lea2014}. 

A comparison between the magnetic field extrapolations for April 15 and 16 shows that the inclination between the two flux systems (blue/orange field lines in Figure~\ref{Fig:T-QSL}) that are involved in the series of flares has changed during these two days. The flux systems are more anti-parallel on April 15, than they are on April 16. This change in the mutual orientation of the field is in agreement with the evolution of the active region. As discussed in Section~\ref{Sect:MF} the negative polarity N3 undergoes a counterclockwise rotation during the two days of observation, while the positive sunspot P1 is seen to rotate clockwise of about 35 degrees. This does not only increase the shear of the magnetic arcade that supports the filament --- as can be deduced from the formation of an S-shaped filament on April 16 (see Figure~\ref{Fig:filament}), but also changes the orientation of the field system making them less favorable to reconnect on April 16 with respect to April 15. 

To discuss the magnetic configuration of the system we used a potential field model. While this model is able to identify the key topological structures of our active region (see Section~\ref{Sect:MFT}), it has some limitations. Firstly, the low lying magnetic field, i.e., the ones that support the filament is definitely in a non-potential state as the clear S-shaped structure of the filament suggests. As a consequence of this extra shear the actual inclination between the magnetic flux system that supports the filament (blue field lines in Figure~\ref{Fig:T-QSL}) and the overlying field (orange field lines) is probably larger, i.e., less anti-parallel, than what predicted by the potential field model. Secondly, we note that the potential field model is an over-relaxed and already fully-reconnected model. As a consequence while it may seem that the discussion of the mutual orientation between the two flux systems applies only at high altitude, in reality the interaction occurs at a lower height. Actually, the blue loops anchored in the moving negative polarity N3 will just rise and collide with P0-N1 orange-type loops that initially have their orientation at low altitude, pushing them up, and reconnecting to form the green loops (Figure~\ref{Fig:T-Overview}). 

During the two days of observations all the (fully and failed) eruptions are initiated around the southern part of the negative polarity N3. The reason for this behavior can be understood from Figure~\ref{Fig:T-QSL}. This is the location of the coronal (quasi-)separator and a perturbation around the (quasi-)separator will initiate magnetic reconnection around its location. As a consequence, the tension of the confining arcade will be  reduced, and the system evolves in the direction of the favorable magnetic pressure gradient, i.e., towards the (quasi-)null-point.   On April 15,  the mutual orientation of the flux systems is favorable for reconnection along the full extension of the filament and we observe a series of full eruptions. On the contrary, on April 16, the mutual orientation is less and less favorable for the reconnection as more and more we move away from the location of the null. The failed eruptions begin from the southern part of the negative polarity N3 that is more prone to reconnection (see Figure~\ref{Fig:T-QSL5}). However, the main part of the filament is located in a region where the  mutual orientation of the flux systems is  less-favorable for magnetic reconnection and the full eruption is eventually inhibited. 

\section{Conclusion}

The aim of this paper was to study the transition from eruptive to confined flares in active region NOAA 12035. This transition occurred between 2014 April 15 and April 16. On April 15, four of the 13 flares observed resulted in a CME, while none of the 13 flares recorded on April 16 resulted in a measurable CME. 

During the two days of observation the filament evolved from being constituted by two separated filaments on April 15 to a single S--shaped filament on April 16.  Contemporaneously to this evolution we observed the presence of significant shear motions that were the results of clockwise/counterclockwise motions of the two magnetic polarities (P1, N3) where the arcade that supports the filament was anchored.  

To study the topology of the active region we performed two potential field extrapolations, one on April 15 and one on April 16, and computed the QSLs. We found that a closed fan-like QSL exists around the location of the filament on both days. The presence of circular, closed fan-QSLs indicate the presence of a (quasi-)separator in the corona. 

The presence of a null-point topology in the corona, the presence of shear motions that reduced the mutual inclination between the two flux systems achieving a configuration less favorable for reconnection, as well as the non significant change in the theoretical stability (with respect to the torus instability scenario) between the two days, leads us to the conclusion that the breakout scenario seems the more probable scenario to describe the observed behavior. The discerning element between fully and failed eruption behavior being determined by the mutual inclination of the flux systems involved in the process, namely the erupting flux and the overlying field.

\begin{acknowledgements}
F.P.Z. is a Fonds Wetenschappelijk Onderzoek (FWO) research fellow (Project No. 1272714N). F.P.Z. acknowledges the AXA Research Fund for contributing to this research. R.J. thanks the DST, Govt. of India for the INSPIRE fellowship. We thank Dr. Wahab Uddin for providing the H-alpha Observations. AIA and HMI data are courtesy of NASA/SDO and the AIA and HMI science teams. The SOHO/LASCO data used here are produced by a consortium of the Naval Research Laboratory (USA), Max-Planck-Institut f\"{u}r Sonnensystemforschung (Germany), Laboratoire d'Astrophysique Marseille (France), and the University of Birmingham (UK). SOHO is a project of international cooperation between ESA and NASA.

\end{acknowledgements}
	
\bibliographystyle{aa} % style aa.bst
   \bibliography{biblio.bib} % your references Yourfile.bib

\begin{thebibliography}{86}
\expandafter\ifx\csname natexlab\endcsname\relax\def\natexlab#1{#1}\fi

\bibitem[{{Alissandrakis}(1981)}]{Ali1981}
{Alissandrakis}, C.~E. 1981, \aap, 100, 197

\bibitem[{{Amari} {et~al.}(2014){Amari}, {Canou}, \& {Aly}}]{Ama2014}
{Amari}, T., {Canou}, A., \& {Aly}, J.-J. 2014, \nat, 514, 465

\bibitem[{{Antiochos} {et~al.}(1999){Antiochos}, {DeVore}, \&
  {Klimchuk}}]{Ant1999}
{Antiochos}, S., {DeVore}, C., \& {Klimchuk}, J. 1999, ApJ, 510, 485

\bibitem[{{Archontis} \& {Hood}(2012)}]{Arc2012}
{Archontis}, V. \& {Hood}, A.~W. 2012, \aap, 537, A62

\bibitem[{{Archontis} \& {T{\"o}r{\"o}k}(2008)}]{Arc2008}
{Archontis}, V. \& {T{\"o}r{\"o}k}, T. 2008, \aap, 492, L35

\bibitem[{{Aulanier}(2014)}]{Aul2014}
{Aulanier}, G. 2014, in IAU Symposium, Vol. 300, IAU Symposium, ed.
  B.~{Schmieder}, J.-M. {Malherbe}, \& S.~T. {Wu}, 184--196

\bibitem[{{Aulanier} {et~al.}(2000){Aulanier}, {DeLuca}, {Antiochos},
  {McMullen}, \& {Golub}}]{Aulanier2000}
{Aulanier}, G., {DeLuca}, E.~E., {Antiochos}, S.~K., {McMullen}, R.~A., \&
  {Golub}, L. 2000, \apj, 540, 1126

\bibitem[{{Aulanier} {et~al.}(2012){Aulanier}, {Janvier}, \&
  {Schmieder}}]{Aul2012}
{Aulanier}, G., {Janvier}, M., \& {Schmieder}, B. 2012, \aap, 543, A110

\bibitem[{{Aulanier} {et~al.}(2006){Aulanier}, {Pariat}, {D{\'e}moulin}, \&
  {DeVore}}]{Aul2006}
{Aulanier}, G., {Pariat}, E., {D{\'e}moulin}, P., \& {DeVore}, C.~R. 2006,
  \solphys, 238, 347

\bibitem[{{Aulanier} {et~al.}(2010){Aulanier}, {T{\"o}r{\"o}k}, {D{\'e}moulin},
  \& {DeLuca}}]{Aul2010}
{Aulanier}, G., {T{\"o}r{\"o}k}, T., {D{\'e}moulin}, P., \& {DeLuca}, E.~E.
  2010, \apj, 708, 314

\bibitem[{{Brueckner} {et~al.}(1995){Brueckner}, {Howard}, {Koomen},
  {Korendyke}, {Michels}, {Moses}, {Socker}, {Dere}, {Lamy}, {Llebaria},
  {Bout}, {Schwenn}, {Simnett}, {Bedford}, \& {Eyles}}]{Bru1995}
{Brueckner}, G.~E., {Howard}, R.~A., {Koomen}, M.~J., {et~al.} 1995, \solphys,
  162, 357

\bibitem[{{Canou} \& {Amari}(2010)}]{Can2010}
{Canou}, A. \& {Amari}, T. 2010, \apj, 715, 1566

\bibitem[{{Carmichael}(1964)}]{Carmichael64}
{Carmichael}, H. 1964, NASA Special Publication, 50, 451

\bibitem[{{Chandra} {et~al.}(2016){Chandra}, {Chen}, {Fulara}, {Srivastava}, \&
  {Uddin}}]{Chandra2016}
{Chandra}, R., {Chen}, P.~F., {Fulara}, A., {Srivastava}, A.~K., \& {Uddin}, W.
  2016, \apj, 822, 106

\bibitem[{{Chandra} {et~al.}(2009){Chandra}, {Schmieder}, {Aulanier}, \&
  {Malherbe}}]{Chandra2009}
{Chandra}, R., {Schmieder}, B., {Aulanier}, G., \& {Malherbe}, J.~M. 2009,
  \solphys, 258, 53

\bibitem[{Chen(2011)}]{Chen2011}
Chen, P.~F. 2011, Living Reviews in Solar Physics, 8

\bibitem[{Chen {et~al.}(2016)Chen, Du, Zhao, Wu, Liu, Wang, Ruan, Feng, \&
  Song}]{Chen2016}
Chen, Y., Du, G., Zhao, D., {et~al.} 2016, The Astrophysical Journal Letters,
  820, L37

\bibitem[{{Dalmasse} {et~al.}(2015){Dalmasse}, {Chandra}, {Schmieder}, \&
  {Aulanier}}]{Dal2015}
{Dalmasse}, K., {Chandra}, R., {Schmieder}, B., \& {Aulanier}, G. 2015, \aap,
  574, A37

\bibitem[{{D{\'e}moulin} \& {Aulanier}(2010)}]{Dem2010}
{D{\'e}moulin}, P. \& {Aulanier}, G. 2010, \apj, 718, 1388

\bibitem[{{D{\'e}moulin} {et~al.}(1996){D{\'e}moulin}, {Priest}, \&
  {Lonie}}]{Dem1996}
{D{\'e}moulin}, P., {Priest}, E.~R., \& {Lonie}, D.~P. 1996, \jgr, 101, 7631

\bibitem[{{Dud{\'{\i}}k} {et~al.}(2014){Dud{\'{\i}}k}, {Janvier}, {Aulanier},
  {Del Zanna}, {Karlick{\'y}}, {Mason}, \& {Schmieder}}]{Dud2014a}
{Dud{\'{\i}}k}, J., {Janvier}, M., {Aulanier}, G., {et~al.} 2014, \apj, 784,
  144

\bibitem[{{Dud{\'{\i}}k} {et~al.}(2016){Dud{\'{\i}}k}, {Polito}, {Janvier},
  {Mulay}, {Karlick{\'y}}, {Aulanier}, {Del Zanna}, {Dzif{\v c}{\'a}kov{\'a}},
  {Mason}, \& {Schmieder}}]{Dud2016}
{Dud{\'{\i}}k}, J., {Polito}, V., {Janvier}, M., {et~al.} 2016, \apj, 823, 41

\bibitem[{{Fan} \& {Gibson}(2007)}]{Fan2007}
{Fan}, Y. \& {Gibson}, S.~E. 2007, \apj, 668, 1232

\bibitem[{{Filippov}(2013)}]{Fil2013}
{Filippov}, B. 2013, \apj, 773, 10

\bibitem[{{Filippov} {et~al.}(2015){Filippov}, {Martsenyuk}, {Srivastava}, \&
  {Uddin}}]{Fil2015}
{Filippov}, B., {Martsenyuk}, O., {Srivastava}, A.~K., \& {Uddin}, W. 2015,
  Journal of Astrophysics and Astronomy [\eprint[arXiv]{1501.02562}]

\bibitem[{{Filippov} \& {Den}(2001)}]{Fil2001}
{Filippov}, B.~P. \& {Den}, O.~G. 2001, \jgr, 106, 25177

\bibitem[{{Forbes}(2010)}]{Forbes2010}
{Forbes}, T. 2010, {Models of coronal mass ejections and flares}, ed.
  {Schrijver, C.~J.~\& Siscoe, G.~L.} (Cambridge University Press), 159

\bibitem[{{Forbes} \& {Isenberg}(1991)}]{For1991}
{Forbes}, T.~G. \& {Isenberg}, P.~A. 1991, \apj, 373, 294

\bibitem[{{Galsgaard} {et~al.}(2007){Galsgaard}, {Archontis},
  {Moreno-Insertis}, \& {Hood}}]{Gal2007}
{Galsgaard}, K., {Archontis}, V., {Moreno-Insertis}, F., \& {Hood}, A.~W. 2007,
  \apj, 666, 516

\bibitem[{{Gibb} {et~al.}(2014){Gibb}, {Mackay}, {Green}, \&
  {Meyer}}]{Gibb2014}
{Gibb}, G.~P.~S., {Mackay}, D.~H., {Green}, L.~M., \& {Meyer}, K.~A. 2014,
  \apj, 782, 71

\bibitem[{{Gibson}(2015)}]{Gib2015}
{Gibson}, S. 2015, in Astrophysics and Space Science Library, Vol. 415, Solar
  Prominences, ed. J.-C. {Vial} \& O.~{Engvold}, 323

\bibitem[{{Gibson} \& {Fan}(2006)}]{Gibson06}
{Gibson}, S.~E. \& {Fan}, Y. 2006, \apjl, 637, L65

\bibitem[{{Gibson} {et~al.}(2010){Gibson}, {Kucera}, {Rastawicki}, {Dove}, {de
  Toma}, {Hao}, {Hill}, {Hudson}, {Marqu{\'e}}, {McIntosh}, {Rachmeler},
  {Reeves}, {Schmieder}, {Schmit}, {Seaton}, {Sterling}, {Tripathi},
  {Williams}, \& {Zhang}}]{Gib2010}
{Gibson}, S.~E., {Kucera}, T.~A., {Rastawicki}, D., {et~al.} 2010, \apj, 724,
  1133

\bibitem[{{Green} {et~al.}(2011){Green}, {Kliem}, \& {Wallace}}]{Gre2011}
{Green}, L.~M., {Kliem}, B., \& {Wallace}, A.~J. 2011, \aap, 526, A2

\bibitem[{{Guo} {et~al.}(2010{\natexlab{a}}){Guo}, {Ding}, {Schmieder}, {Li},
  {T{\"o}r{\"o}k}, \& {Wiegelmann}}]{Guo2010a}
{Guo}, Y., {Ding}, M.~D., {Schmieder}, B., {et~al.} 2010{\natexlab{a}}, \apjl,
  725, L38

\bibitem[{{Guo} {et~al.}(2010{\natexlab{b}}){Guo}, {Schmieder}, {D{\'e}moulin},
  {Wiegelmann}, {Aulanier}, {T{\"o}r{\"o}k}, \& {Bommier}}]{Guo2010b}
{Guo}, Y., {Schmieder}, B., {D{\'e}moulin}, P., {et~al.} 2010{\natexlab{b}},
  \apj, 714, 343

\bibitem[{{Harra} {et~al.}(2016){Harra}, {Schrijver}, {Janvier}, {Toriumi},
  {Hudson}, {Matthews}, {Woods}, {Hara}, {Guedel}, {Kowalski}, {Osten},
  {Kusano}, \& {Lueftinger}}]{Har2016}
{Harra}, L.~K., {Schrijver}, C.~J., {Janvier}, M., {et~al.} 2016, \solphys,
  291, 1761

\bibitem[{{Hirayama}(1974)}]{Hirayama74}
{Hirayama}, T. 1974, \solphys, 34, 323

\bibitem[{{Inoue} {et~al.}(2015){Inoue}, {Hayashi}, {Magara}, {Choe}, \&
  {Park}}]{Ino2015}
{Inoue}, S., {Hayashi}, K., {Magara}, T., {Choe}, G.~S., \& {Park}, Y.~D. 2015,
  \apj, 803, 73

\bibitem[{{Isenberg} \& {Forbes}(2007)}]{Ise2007}
{Isenberg}, P.~A. \& {Forbes}, T.~G. 2007, \apj, 670, 1453

\bibitem[{{Janvier} {et~al.}(2015){Janvier}, {Aulanier}, \&
  {D{\'e}moulin}}]{Jan2015}
{Janvier}, M., {Aulanier}, G., \& {D{\'e}moulin}, P. 2015, \solphys, 290, 3425

\bibitem[{{Janvier} {et~al.}(2013){Janvier}, {Aulanier}, {Pariat}, \&
  {D{\'e}moulin}}]{Jan2013}
{Janvier}, M., {Aulanier}, G., {Pariat}, E., \& {D{\'e}moulin}, P. 2013, \aap,
  555, A77

\bibitem[{{Jiang} {et~al.}(2014){Jiang}, {Wu}, {Feng}, \& {Hu}}]{Jiang2014}
{Jiang}, C., {Wu}, S.~T., {Feng}, X., \& {Hu}, Q. 2014, \apjl, 786, L16

\bibitem[{{Jing} {et~al.}(2010){Jing}, {Yuan}, {Wiegelmann}, {Xu}, {Liu}, \&
  {Wang}}]{Jing2010}
{Jing}, J., {Yuan}, Y., {Wiegelmann}, T., {et~al.} 2010, \apjl, 719, L56

\bibitem[{{Karpen} {et~al.}(2012){Karpen}, {Antiochos}, \&
  {DeVore}}]{Karpen2012}
{Karpen}, J.~T., {Antiochos}, S.~K., \& {DeVore}, C.~R. 2012, \apj, 760, 81

\bibitem[{{Kliem} {et~al.}(2014{\natexlab{a}}){Kliem}, {Lin}, {Forbes},
  {Priest}, \& {T{\"o}r{\"o}k}}]{Kliem2014}
{Kliem}, B., {Lin}, J., {Forbes}, T.~G., {Priest}, E.~R., \& {T{\"o}r{\"o}k},
  T. 2014{\natexlab{a}}, \apj, 789, 46

\bibitem[{{Kliem} {et~al.}(2013){Kliem}, {Su}, {van Ballegooijen}, \&
  {DeLuca}}]{Kliem2013}
{Kliem}, B., {Su}, Y.~N., {van Ballegooijen}, A.~A., \& {DeLuca}, E.~E. 2013,
  \apj, 779, 129

\bibitem[{{Kliem} \& {T{\"o}r{\"o}k}(2006)}]{Kliem2006}
{Kliem}, B. \& {T{\"o}r{\"o}k}, T. 2006, Physical Review Letters, 96, 255002

\bibitem[{{Kliem} {et~al.}(2014{\natexlab{b}}){Kliem}, {T{\"o}r{\"o}k},
  {Titov}, {Lionello}, {Linker}, {Liu}, {Liu}, \& {Wang}}]{Kliem14}
{Kliem}, B., {T{\"o}r{\"o}k}, T., {Titov}, V.~S., {et~al.} 2014{\natexlab{b}},
  \apj, 792, 107

\bibitem[{{Kopp} \& {Pneuman}(1976)}]{Kopp76}
{Kopp}, R.~A. \& {Pneuman}, G.~W. 1976, \solphys, 50, 85

\bibitem[{{Leake} {et~al.}(2014){Leake}, {Linton}, \& {Antiochos}}]{Lea2014}
{Leake}, J.~E., {Linton}, M.~G., \& {Antiochos}, S.~K. 2014, \apj, 787, 46

\bibitem[{{Leake} {et~al.}(2013){Leake}, {Linton}, \&
  {T{\"o}r{\"o}k}}]{Lea2013}
{Leake}, J.~E., {Linton}, M.~G., \& {T{\"o}r{\"o}k}, T. 2013, \apj, 778, 99

\bibitem[{{Lemen} {et~al.}(2012){Lemen}, {Title}, {Akin}, {Boerner}, {Chou},
  {Drake}, {Duncan}, {Edwards}, {Friedlaender}, {Heyman}, {Hurlburt}, {Katz},
  {Kushner}, {Levay}, {Lindgren}, {Mathur}, {McFeaters}, {Mitchell}, {Rehse},
  {Schrijver}, {Springer}, {Stern}, {Tarbell}, {Wuelser}, {Wolfson}, {Yanari},
  {Bookbinder}, {Cheimets}, {Caldwell}, {Deluca}, {Gates}, {Golub}, {Park},
  {Podgorski}, {Bush}, {Scherrer}, {Gummin}, {Smith}, {Auker}, {Jerram},
  {Pool}, {Soufli}, {Windt}, {Beardsley}, {Clapp}, {Lang}, \& {Waltham}}]{AIA}
{Lemen}, J.~R., {Title}, A.~M., {Akin}, D.~J., {et~al.} 2012, \solphys, 275, 17

\bibitem[{{Liu} {et~al.}(2008){Liu}, {Gilbert}, {Alexander}, \& {Su}}]{Liu08}
{Liu}, R., {Gilbert}, H.~R., {Alexander}, D., \& {Su}, Y. 2008, \apj, 680, 1508

\bibitem[{{Liu} {et~al.}(2012){Liu}, {Kliem}, {T{\"o}r{\"o}k}, {Liu}, {Titov},
  {Lionello}, {Linker}, \& {Wang}}]{Liu12}
{Liu}, R., {Kliem}, B., {T{\"o}r{\"o}k}, T., {et~al.} 2012, \apj, 756, 59

\bibitem[{{Lynch} {et~al.}(2008){Lynch}, {Antiochos}, {DeVore}, {Luhmann}, \&
  {Zurbuchen}}]{Lyn2008}
{Lynch}, B.~J., {Antiochos}, S.~K., {DeVore}, C.~R., {Luhmann}, J.~G., \&
  {Zurbuchen}, T.~H. 2008, ApJ, 683, 1192

\bibitem[{{Mandrini} {et~al.}(2006){Mandrini}, {Demoulin}, {Schmieder},
  {Deluca}, {Pariat}, \& {Uddin}}]{Man2006}
{Mandrini}, C.~H., {Demoulin}, P., {Schmieder}, B., {et~al.} 2006, \solphys,
  238, 293

\bibitem[{{Masson} {et~al.}(2009){Masson}, {Pariat}, {Aulanier}, \&
  {Schrijver}}]{Mas2009}
{Masson}, S., {Pariat}, E., {Aulanier}, G., \& {Schrijver}, C.~J. 2009, \apj,
  700, 559

\bibitem[{{McCauley} {et~al.}(2015){McCauley}, {Su}, {Schanche}, {Evans}, {Su},
  {McKillop}, \& {Reeves}}]{McC2015}
{McCauley}, P.~I., {Su}, Y.~N., {Schanche}, N., {et~al.} 2015, \solphys, 290,
  1703

\bibitem[{{Pariat} \& {D{\'e}moulin}(2012)}]{Par2012}
{Pariat}, E. \& {D{\'e}moulin}, P. 2012, \aap, 541, A78

\bibitem[{{Pesnell} {et~al.}(2012){Pesnell}, {Thompson}, \& {Chamberlin}}]{SDO}
{Pesnell}, W.~D., {Thompson}, B.~J., \& {Chamberlin}, P.~C. 2012, \solphys,
  275, 3

\bibitem[{{Rachmeler} {et~al.}(2013){Rachmeler}, {Gibson}, {Dove}, {DeVore}, \&
  {Fan}}]{Rac2013}
{Rachmeler}, L.~A., {Gibson}, S.~E., {Dove}, J.~B., {DeVore}, C.~R., \& {Fan},
  Y. 2013, \solphys, 288, 617

\bibitem[{{Romano} {et~al.}(2014){Romano}, {Zuccarello}, {Guglielmino}, \&
  {Zuccarello}}]{Rom2014}
{Romano}, P., {Zuccarello}, F.~P., {Guglielmino}, S.~L., \& {Zuccarello}, F.
  2014, \apj, 794, 118

\bibitem[{{Savcheva} {et~al.}(2016){Savcheva}, {Pariat}, {McKillop},
  {McCauley}, {Hanson}, {Su}, \& {DeLuca}}]{Sav2016}
{Savcheva}, A., {Pariat}, E., {McKillop}, S., {et~al.} 2016, \apj, 817, 43

\bibitem[{{Savcheva} {et~al.}(2015){Savcheva}, {Pariat}, {McKillop},
  {McCauley}, {Hanson}, {Su}, {Werner}, \& {DeLuca}}]{Sav2015}
{Savcheva}, A., {Pariat}, E., {McKillop}, S., {et~al.} 2015, \apj, 810, 96

\bibitem[{{Savcheva} {et~al.}(2012){Savcheva}, {Green}, {van Ballegooijen}, \&
  {DeLuca}}]{Sav2012}
{Savcheva}, A.~S., {Green}, L.~M., {van Ballegooijen}, A.~A., \& {DeLuca},
  E.~E. 2012, \apj, 759, 105

\bibitem[{{Schmieder} {et~al.}(1997){Schmieder}, {Aulanier}, {Demoulin}, {van
  Driel-Gesztelyi}, {Roudier}, {Nitta}, \& {Cauzzi}}]{Sch1997}
{Schmieder}, B., {Aulanier}, G., {Demoulin}, P., {et~al.} 1997, \aap, 325, 1213

\bibitem[{{Schmieder} {et~al.}(2015){Schmieder}, {Aulanier}, \& {Vr{\v
  s}nak}}]{Sch2015}
{Schmieder}, B., {Aulanier}, G., \& {Vr{\v s}nak}, B. 2015, \solphys, 290, 3457

\bibitem[{{Schou} {et~al.}(2012){Schou}, {Scherrer}, {Bush}, {Wachter},
  {Couvidat}, {Rabello-Soares}, {Bogart}, {Hoeksema}, {Liu}, {Duvall}, {Akin},
  {Allard}, {Miles}, {Rairden}, {Shine}, {Tarbell}, {Title}, {Wolfson},
  {Elmore}, {Norton}, \& {Tomczyk}}]{HMI}
{Schou}, J., {Scherrer}, P.~H., {Bush}, R.~I., {et~al.} 2012, \solphys, 275,
  229

\bibitem[{{Sturrock}(1966)}]{Sturrock66}
{Sturrock}, P.~A. 1966, \nat, 211, 695

\bibitem[{{Sun} {et~al.}(2015){Sun}, {Bobra}, {Hoeksema}, {Liu}, {Li}, {Shen},
  {Couvidat}, {Norton}, \& {Fisher}}]{Sun2015}
{Sun}, X., {Bobra}, M.~G., {Hoeksema}, J.~T., {et~al.} 2015, \apjl, 804, L28

\bibitem[{{Sun} {et~al.}(2013){Sun}, {Hoeksema}, {Liu}, {Aulanier}, {Su},
  {Hannah}, \& {Hock}}]{Sun2013}
{Sun}, X., {Hoeksema}, J.~T., {Liu}, Y., {et~al.} 2013, \apj, 778, 139

\bibitem[{{Thalmann} {et~al.}(2015){Thalmann}, {Su}, {Temmer}, \&
  {Veronig}}]{Tha2015}
{Thalmann}, J.~K., {Su}, Y., {Temmer}, M., \& {Veronig}, A.~M. 2015, \apjl,
  801, L23

\bibitem[{{Titov} {et~al.}(2002){Titov}, {Hornig}, \& {D{\'e}moulin}}]{Tit2002}
{Titov}, V.~S., {Hornig}, G., \& {D{\'e}moulin}, P. 2002, Journal of
  Geophysical Research (Space Physics), 107, 1164

\bibitem[{{T{\"o}r{\"o}k} \& {Kliem}(2005)}]{Tor2005}
{T{\"o}r{\"o}k}, T. \& {Kliem}, B. 2005, \apjl, 630, L97

\bibitem[{{T{\"o}r{\"o}k} \& {Kliem}(2007)}]{Tor2007}
{T{\"o}r{\"o}k}, T. \& {Kliem}, B. 2007, Astronomische Nachrichten, 328, 743

\bibitem[{{Tripathi} {et~al.}(2013){Tripathi}, {Reeves}, {Gibson},
  {Srivastava}, \& {Joshi}}]{Tripathi13}
{Tripathi}, D., {Reeves}, K.~K., {Gibson}, S.~E., {Srivastava}, A., \& {Joshi},
  N.~C. 2013, \apj, 778, 142

\bibitem[{{Yashiro} {et~al.}(2005){Yashiro}, {Gopalswamy}, {Akiyama},
  {Michalek}, \& {Howard}}]{Yas2005}
{Yashiro}, S., {Gopalswamy}, N., {Akiyama}, S., {Michalek}, G., \& {Howard},
  R.~A. 2005, Journal of Geophysical Research (Space Physics), 110, A12S05

\bibitem[{{Zhao} {et~al.}(2016){Zhao}, {Gilchrist}, {Aulanier}, {Schmieder},
  {Pariat}, \& {Li}}]{Zhao2016}
{Zhao}, J., {Gilchrist}, S.~A., {Aulanier}, G., {et~al.} 2016, \apj, 823, 62

\bibitem[{{Zhao} {et~al.}(2014){Zhao}, {Li}, {Pariat}, {Schmieder}, {Guo}, \&
  {Wiegelmann}}]{Zhao2014}
{Zhao}, J., {Li}, H., {Pariat}, E., {et~al.} 2014, \apj, 787, 88

\bibitem[{{Zhu} \& {Alexander}(2014)}]{Zhang14}
{Zhu}, C. \& {Alexander}, D. 2014, \solphys, 289, 279

\bibitem[{{Zuccarello} {et~al.}(2015){Zuccarello}, {Aulanier}, \&
  {Gilchrist}}]{Zuc2015}
{Zuccarello}, F.~P., {Aulanier}, G., \& {Gilchrist}, S.~A. 2015, \apj, 814, 126

\bibitem[{{Zuccarello} {et~al.}(2016){Zuccarello}, {Aulanier}, \&
  {Gilchrist}}]{Zuc2016}
{Zuccarello}, F.~P., {Aulanier}, G., \& {Gilchrist}, S.~A. 2016, \apjl, 821,
  L23

\bibitem[{{Zuccarello} {et~al.}(2009){Zuccarello}, {Jacobs}, {Soenen},
  {Poedts}, {van der Holst}, \& {Zuccarello}}]{Zuc2009}
{Zuccarello}, F.~P., {Jacobs}, C., {Soenen}, A., {et~al.} 2009, \aap, 507, 441

\bibitem[{{Zuccarello} {et~al.}(2014){Zuccarello}, {Seaton}, {Mierla},
  {Poedts}, {Rachmeler}, {Romano}, \& {Zuccarello}}]{Zuc2014}
{Zuccarello}, F.~P., {Seaton}, D.~B., {Mierla}, M., {et~al.} 2014, \apj, 785,
  88

\bibitem[{{Zuccarello} {et~al.}(2008){Zuccarello}, {Soenen}, {Poedts},
  {Zuccarello}, \& {Jacobs}}]{Zuc2008}
{Zuccarello}, F.~P., {Soenen}, A., {Poedts}, S., {Zuccarello}, F., \& {Jacobs},
  C. 2008, ApJ, 689, L157

\end{thebibliography}
\end{document}